\documentclass[11pt]{article}

\usepackage{epsfig}  

\usepackage{amsmath}
\usepackage{amssymb}
\usepackage{eepic}
\usepackage{epic}

\newcommand{\D}{\displaystyle}
\newcommand{\bc}{\mathbb{C}}
\newcommand{\br}{\mathbb{R}}
\newcommand{\bp}{\mathbb{P}}
\newcommand{\bpxp}{\mathbb{P}^1\times\mathbb{P}^1}
\newcommand{\bz}{\mathbb{Z}}

\usepackage{amsthm}
\theoremstyle{definition}
\newtheorem{theorem}{Theorem}[section]
\newtheorem{lemma}[theorem]{Lemma}
\newtheorem{proposition}[theorem]{Proposition}

\newtheorem{example}[theorem]{Example}

\begin{document}

\title
{\vskip -70pt
\vskip 2cm
{\huge {Spectral curves and the mass of \\ 
hyperbolic monopoles}}\\  
\vspace{1mm}
}

\author{
{\Large \bf Paul Norbury}
\thanks{e-mail: {\tt pnorbury@ms.unimelb.edu.au}}\\[1mm]
{\normalsize \sl Department of Mathematics and Statistics}\\
{\normalsize \sl University of Melbourne, 3010 Australia}\\[4mm]
{\Large \bf Nuno M. Rom\~ao}
\thanks{e-mail: {\tt nromao@maths.adelaide.edu.au}}\\[1mm]
{\normalsize \sl School of Mathematical Sciences}\\
{\normalsize \sl University of Adelaide, 5005 Australia}\\[3mm]
}

\date{December 2005}

\maketitle

\begin{abstract}

\noindent The moduli spaces of hyperbolic monopoles are naturally
fibred by the monopole mass, and this leads to a nontrivial mass
dependence of the holomorphic data (spectral curves, rational maps,
holomorphic spheres)
associated to hyperbolic multi-monopoles.  In
this paper, we obtain an explicit description of this dependence for
general hyperbolic monopoles of magnetic charge two. In addition,
we show how to compute the monopole mass of higher charge spectral
curves with tetrahedral and octahedral symmetries. Spectral curves
of euclidean monopoles are recovered from our results via an infinite-mass
limit.

\end{abstract}

\vspace{3mm}


\vspace{2mm}


\section{Introduction}

Magnetic monopoles are paradigmatic examples of topological solitons
in gauged field theories.  Once a metric on $\mathbb{R}^3$ has been
fixed, a monopole is defined as a pair $({\rm d}_A,\Phi)$ satisfying the
Bogomol'ny\u\i\ equation
\begin{equation}\label{Bog}
    B_A= *{\rm d}_{A}\Phi
\end{equation}
modulo gauge equivalence.  Here, $B_A$ is the curvature of the
${\rm SU}(2)$-connection ${\rm d}_A$ on ${\rm End}\, {\cal E}$, where ${\cal E} \rightarrow \br^3$
is the trivial vector bundle associated to the defining representation, and $\Phi$
(the Higgs field) is a section of ${\rm End}\, {\cal E}$ with constant and
nonzero norm at infinity,
\begin{equation}\label{mass}
    \lim_{|\mathbf{x}|\rightarrow \infty} ||\Phi(\mathbf{x})|| = m>0.
\end{equation}
In the radial compactification of $\mathbb{R}^{3}$, the boundary
condition (\ref{mass}) provides a map 
\[
\Phi|_{S^{2}_{\infty}}:S^2\rightarrow S^{2}
\] 
from the 2-sphere at infinity to the 2-sphere of radius $m$ centred at
the origin of $\mathfrak{su}(2) \cong \mathbb{R}^{3}$, whose degree
$k$ is interpreted as the magnetic charge of the monopole.  The
Bogomol'ny\u\i\ equation (\ref{Bog}) turns out to be quite tractable
when the metric has constant curvature, for then twistor methods can
be used to characterise the solutions in terms of objects from complex
geometry (holomorphic bundles, spectral curves, rational maps).  There
is an extensive literature focused on the case where the metric is
euclidean.  Then the space of all monopoles with a given topology
carries a natural $L^2$ metric which is hyperk\"ahler, and its
geodesic flow can be interpreted physically as slow motion in the
Yang--Mills--Higgs model in
$\mathbb{R}^{1,3}$~\cite{Manrk,AtiHitMM,ManSut}.  The algebraic
topology of these moduli spaces captures the spectrum of bound states
at the quantum level, and it has been shown to be consistent with
certain S-duality conjectures in quantum field
theory~\cite{Sen,SegSel}.

Monopoles in hyperbolic space, where the metric in polar coordinates
is taken to be 
\[ 
ds^2=dr^2 + R^4 \sinh^{2}\left(\frac{r}{R^2}\right)
\left(d\theta^2 + \sin^2 \theta \, d\phi^{2} \right),
\] 
were first sudied by Atiyah~\cite{AtiMMH}, and they were soon
perceived as being quite distinct from euclidean monopoles.  They have
a remarkably rich structure, which somewhat degenerates when the
euclidean limit $R \rightarrow \infty$ is taken.  One interesting fact
about hyperbolic monopoles is that they are completely determined by
the value of the connection at infinity~\cite{NorBA,MurNorSinHS},
which reduces to a ${\rm U}(1)$-connection on a 2-sphere; this can be
regarded as a classical manifestation of the AdS/CFT correspondence~\cite{Mal}. 
The connection at infinity also plays a r\^ole in a characterisation of
hyperbolic monopoles through holomorphic spheres in projective
spaces~\cite{MurNorSinHS}, which is not available in the euclidean
case.

A fundamental feature of hyperbolic monopoles that makes them
different from their euclidean counterparts is that they come with
different values of the mass $m \in \; ]0,\infty[$ defined by equation
(\ref{mass}).  This parameter is related to the radius of curvature
$R$ of hyperbolic space: a rescaling of the radius by $R\mapsto
\lambda R$ maps monopoles of mass $m$ to monopoles of mass $m/\lambda
$.  (In the euclidean case, the radius of curvature is infinite and
this means that euclidean monopoles of different masses can be
identified.)  Fixing the radius of hyperbolic space, the mass is a
physical parameter and naturally appears as one of the moduli for the
solutions of the Bogomol'ny\u\i\ equations.  The objects obtained in
the limit of mass zero (sometimes called ``nullarons''\footnote{This
term is due to Michael Murray.}) are somewhat special and simpler to
study; their twistor data have been pointed out to be directly related
to complex curves arising in a construction of solutions to the
Yang--Baxter equation for the chiral Potts
model~\cite{AtiMurMYB,AtiMMYB}.  By the same scaling argument,
hyperbolic monopoles with infinite mass can be identified with
euclidean monopoles~\cite{JarNorZI}.

One of the most basic questions one can ask about the moduli space of
hyperbolic monopoles with a given topology is how this space is
foliated by the mass.  The answer to this question is trivial for
monopoles of magnetic charge one.
For charge two, a partial answer has been given in
reference~\cite{MurNorSinHS}, where a distribution tangent to the
leaves of this foliation was constructed.  In this paper, we will give
a much better characterisation of the mass of a two-monopole.  We
focus on the description of hyperbolic monopoles in terms of spectral
curves, and parametrise spectral curves of 2-monopoles explicitly in
terms of the mass parameter.  In this way, we obtain a complete
characterisation of all the moduli of two-monopoles in hyperbolic
space.  Our methods will apply to calculate the monopole mass in a
continuous family of spectral curves of any charge, but are most
effective in situations analogous to the charge two case, where we 
can take advantage of the fact that each spectral
curve is elliptic. 
In higher charge, one can try to obtain spectral curves that are Galois 
covers of an elliptic curve by imposing platonic symmetries on the monopole
fields, although such symmetric curves will not always correspond 
automatically to smooth solutions of the Bogomol'ny\u\i\ equation 
(\ref{Bog}). We shall undertake a systematic study of this class of curves,
and use their symmetry to reduce the computation of the mass 
dependence, once again, to a calculation on elliptic curves.
Thus we shall provide the first explicit examples of spectral curves of hyperbolic monopoles of arbitrary mass $m>0$ and charges two, three and four
beyond the (rather degenerate) axially symmetric case treated in
reference~\cite{MurNorSinHS}.
We also investigate some limiting cases of our constructions; in particular,
we will show how one can study the infinite-mass limit to recover
spectral curves of euclidean monopoles that have featured in the
existing literature~\cite{Hur, HitManMur, HouSutTC}.

\section{Spectral curves of hyperbolic monopoles}

The setup for twistor theory of hyperbolic space $H^3$ is the 
correspondence
\begin{equation} \label{corr}
    \begin{array}{rcl}
	&{\cal C}&\\
	^{\mu}\swarrow &&\searrow^{\nu} \\[6pt]
	H^3\quad && \quad \; Z\\
    \end{array}
\end{equation}
that we now explain.  We consider the model of $H^3$ as the upper
half-plane $x_{3}>0$ in $\mathbb{R}^{3}$, which we parametrise using
cartesian coordinates $(x_1,x_2,x_3)$.  It is useful to compactify
$H^3$ by adding a boundary $\partial H^{3}\cong \mathbb{P}^{1}$ (the
2-sphere $S^{2}_{\infty}$ at infinity) obtained as a one-point
compactification of the 2-plane of equation $x_{3}=0$.  We use
$z=x_{1}+ix_{2}$ as a stereographic coordinate on this Riemann sphere
($z=\frac{z_1}{z_0}$ in terms of homogeneous coordinates for this
$\mathbb{P}^1$, and $z=\infty$ denotes the point at infinity).  The
geodesics on $H^3$ are the half-circles in $\mathbb{R}^{3}$ lying on
planes perpendicular to and centred at points of $x_{3}=0$, together
with the half-lines perpendicular to this plane, and are uniquely
determined by a pair of points of intersection with $\partial H^3$. 
So the space of all oriented geodesics (the twistor space of $H^3$) is
the complex surface 
\[ 
Z=\mathbb{P}^{1}\times \mathbb{P}^1-\mathbb{P}^{1}_{\bar{\Delta}},
\] 
where
\[
\mathbb{P}^{1}_{\bar{\Delta}}:=\{(w,z)\in\mathbb{P}^{1}\times
\mathbb{P}^{1}: \hat{w}\neq z \}
\] 
is the antidiagonal; $(w,z)$ denotes the oriented geodesic starting at
the antipodal point $\hat{w}:=-1/\bar{w}$ of $w$ and ending at $z$. 
The correspondence space $\cal C$ in (\ref{corr}) is the subset of
$H^3\times Z$ defined by the incidence relation
\begin{equation}\label{incidence}
    (x_1,x_2,x_3) \quad \mbox{lies on}\quad (w,z),
\end{equation}
while the maps $\mu$ and $\nu$ are the natural projections.  Thus $\mu
\circ \nu^{-1} (w,z)$ is the geodesic of $H^3$ corresponding to the
oriented geodesic $(w,z)$, while $\nu \circ
\mu^{-1}(x_{1},x_{2},x_{3})$ is called the star at
$(x_{1},x_{2},x_{3})$ and is the set of all oriented geodesics through
this point of $H^3$.  There are two natural maps $\varepsilon_{\pm}:
Z\rightarrow \mathbb{P}^1$ 
\[
\varepsilon_{-}(w,z)=\hat{w},\qquad \varepsilon_{+}(w,z)=z
\] 
giving the endpoints of an oriented geodesic.

\begin{lemma}\label{th:stars}
    The star at $(x_1,x_2,x_3)$ is the projective line in 
    $Z$ given by the equation
    \begin{equation}\label{star}
	(x_1-ix_2)wz-(x_1^2 + x_2^2+x_3^2)w+z-(x_1+ix_2)=0.
    \end{equation}
\end{lemma}
\begin{proof}
    We observe that if $(w,z)$ is in the star at $(0,0,x_3)$, then $z$
    and $w$ must have the same argument, and Pythagoras' theorem gives
    \begin{equation}\label{star0}
	\left|\frac{\hat{w}+z}{2}\right|^2 = x_{3}^{2}+\left|
	\frac{\hat{w}-z}{2}\right|^2 \qquad \Leftrightarrow \qquad
	{z}=x_3^2 \,w,
    \end{equation}
    which is the equation for the star at $(0,0,x_3)$.  This star is
    related to the star at $(x_1,x_2,x_3)$ by a translation which maps
    an oriented geodesic $(w',z')$ satisfying (\ref{star0}) to an
    oriented geodesic $(w,z)$ through $(x_1,x_2,x_3)$, where
    $z'=z-(x_1+ix_2)$ and 
    \[ 
    w'=\frac{w}{1+(x_1-ix_2)w}.
    \]
    Substituting this in equation (\ref{star0}), we obtain
    (\ref{star}).  It is clear that (\ref{star0}) defines a projective
    line, and so the same must be true for the general star
    (\ref{star}).
\end{proof}

Note that the statements (\ref{incidence}) and (\ref{star}) are
equivalent. 
A star (\ref{star}) is invariant under the map
\begin{equation}\label{reality}
    \sigma: (w,z)\mapsto (\hat{z},\hat{w})
\end{equation}
reversing the orientation of oriented geodesics.  This map is
antiholomorphic and squares to the identity, so it is a reality
structure for the complex surface $Z$; it has no fixed points.  A set
invariant under $\sigma$ is said to be real.

Twistor theory consists of interpreting analytic objects in physical
space $H^3$ in terms of algebraic objects in twistor space $Z$, and
vice-versa.  In our context, the correspondence will relate hyperbolic
monopoles $({\rm d}_{A},\Phi)$ on $H^3$ to their spectral curves, which are
complex curves $\Sigma \subset Z$ satisfying certain conditions.  We
now review briefly how they arise and refer the reader to references
\cite{MurSinSC} and \cite{MurNorSinHS} for further details.

The twistor correspondence for the Bogomol'ny\u\i\ equation
(\ref{Bog}) makes use of the family of operators
\[
\mathfrak{H}: Z \longrightarrow \bigoplus_{(w,z) \in Z} {\rm
End}\,H^0(\mu\circ \nu^{-1}(w,z),{\cal E}).
\] 
introduced by Hitchin (for euclidean monopoles) 
in~\cite{HitMG}, and defined by
\[
\mathfrak{H}(w,z):=\left({\rm d}_{A}-i\Phi\right)|_{\mu \circ
\nu^{-1}(w,z)}.
\]
The Bogomol'ny\u\i\ equation (\ref{Bog}) turns out to be the
integrability condition for the holomorphic structure $(\mu \circ
\nu^{-1})^{*}\bar{\partial}_{A}$ on the bundle $\ker \mathfrak{H}
\rightarrow Z$ induced by the connection ${\rm d}_A$.  Thus, for a solution
$({\rm d}_A,\Phi)$ of (\ref{Bog}), $\ker \mathfrak{H}$ is a holomorphic
complex bundle of rank two, and it has two distinguished line
subbundles $L^{\pm}$ with fibres
\[
L^{\pm}_{(w,z)}=\left\{ s \in \ker \mathfrak{H}(w,z): 
\lim_{x\rightarrow \varepsilon_{\pm}(w,z)} s(x) = 0 \right\}.
\]
The spectral curve $\Sigma \subset Z$ of the monopole $({\rm d}_A,\Phi)$ is
defined as the support of the cokernel of a morphism of coherent
sheaves on $Z$ given by the composition
\[
{\cal O} (L^{-}) \longrightarrow {\cal O} (\ker \mathfrak{H})
\longrightarrow {\cal O} (\ker \mathfrak{H}/L^{+}),
\]
which is the same as saying
\[
\Sigma:=\left\{ (w,z)\in Z: L^{+}_{(w,z)}=L^{-}_{(w,z)}  \right\}.
\]

Holomorphic line bundles on $Z$ can be constructed by purely algebraic
means.  They are obtained as tensor products of bundles pulled back
from each factor of $\mathbb{P}^1\times\mathbb{P}^1$,
\[
{\cal O}_{Z}(k_1,k_2):= \hat{\varepsilon}_{-}^{*}{\cal
O}_{\mathbb{P}^1}(k_1) \otimes {\varepsilon_{+}^{*}}{\cal
O}_{\mathbb{P}^1}(k_2), \qquad k_1,k_2\in \mathbb{Z},
\]
and complex powers $L^{\lambda}$ ($\lambda \in \mathbb{C}$) of the
topologically trivial line bundle
\[
L:={\cal O}_Z (1,-1).
\]
The bundles $L^\lambda$ do not extend to $\mathbb{P}^1 \times
\mathbb{P}^1 \supset Z$ unless $\lambda \in \mathbb{Z}$.  We shall
write
\begin{equation} \label{linebundles}
    L^{\lambda}(k_1,k_2):=L^{\lambda}\otimes {\cal O}_{Z}(k_1,k_2).
\end{equation}
These line bundles can be constructed as follows. Introduce
\begin{equation}\label{U1U2}
    \begin{array}{c}
	U_1:=\{ (w,z) \in Z : w\neq \infty, z\neq \infty \},\\
	U_2:=\{ (w,z) \in Z : w\neq 0, z\neq 0 \}.;
    \end{array}
\end{equation}
since $(0,\infty), (\infty,0) \in \mathbb{P}^{1}_{\bar{\Delta}}$,
these contractible sets provide an open cover of the twistor space $Z$.  It
can be taken as a trivialising cover for the line bundle
(\ref{linebundles}), with transition function $g_{12}:U_1\ \cap U_2
\rightarrow \mathbb{C}^{*}$ given by
\begin{equation}\label{trans}
    g_{12}(w,z)=w^{k_1 + \lambda}z^{k_2 - \lambda}.
\end{equation}

The connection between the holomorphic theory of the Bogomol'ny\u\i\
equation above and algebraic geometry is provided by the
identifications
\[
L^{+}=L^{m}(0,-k)\quad \mbox{and} \quad L^{-}=L^{-m}(-k,0)
\]
for a hyperbolic monopole of charge $k$ and mass $m$.  This result is
obtained from an analysis of the asymptotics of the fields
$({\rm d}_A,\Phi)$ near the boundary $\partial H^3$ \cite{MurSinSC}.
Spectral curves of hyperbolic monopoles satisfy the following three
conditions:

\begin{enumerate}
    \item 
    $\Sigma\subset Z$ is a real curve in the linear system $|{\cal
    O}_{Z}(k,k)|$;
    \item 
    $\left.  L^{2m+k} \right|_{\Sigma} \cong {\cal O}_{\Sigma}$;
    \item 
    $H^{0}(\Sigma,L^{s}(k-2,0))=0$ for all $0<s<2m+2$.
\end{enumerate}

In the broader context of integrable systems, these conditions lead to
an ODE of Lax type. It is believed that they are also sufficient to
guarantee the existence of a hyperbolic monopole to which $\Sigma$ is
associated. Although sufficiency has not yet been proven, for the purposes
of this paper we will use the terms hyperbolic monopole and spectral
curve to refer to complex curves in $Z$ satisfying the conditions above.

Using the coordinates $w,z$ for $Z$, one can write a polynomial equation for $\Sigma$ as
\[
\psi(w,z)=(-w)^k\langle q(\hat{w}),q(z)\rangle 
\]
where $\langle\cdot, \cdot\rangle$ is the standard inner product in $\bc^{k+1}$ and $q:\bc \rightarrow \bc^{k+1}$ is defined in terms of $k+1$ vectors $v_j \in \bc^{k+1}$ by
\[
q(z)=\sum_{j=0}^{k}\sqrt{{k \choose j}}z^j v_j;
\]
in fact, $q$ induces a holomorphic map $\bp^1 \rightarrow \bp^k$ that
also characterises the monopole up to the action of ${\rm PU}(k+1)$ on the target~\cite{MurNorSinHS}.

The conditions 1.--3.\ above yield a model for the moduli space of
hyperbolic $k$-monopoles ${\cal M}_{k}$ in terms of spectral curves. 
A version of this was used in reference~\cite{MurNorSinHS} to give a
description of ${\cal M}_2$.
The group of (direct) isometries of $H^{3}$ can be identified with
${\rm PSL}_2\bc$, and it acts on ${\cal M}_k$. There is a moment map
associated to this action, which can be used to define a centre of
a monopole: spectral curves in the zero-set of the moment map
correspond to centred monopoles and satisfy~\cite{MurNorSinHS}
\[
\sum_{j=0}^k(2j-k)||v_j||^2=0, \qquad\sum_{j=0}^{k-1}\sqrt{(j+1)(k-j)}
\langle v_j, v_{j+1}\rangle=0.
\]

\section{The mass of hyperbolic 2-monopoles}

Our aim in this section is to obtain an explicit equation for the
spectral curve $\Sigma\subset Z$ of a hyperbolic 2-monopole in terms
of its mass.  We shall first deal with the case where $\Sigma$ is
smooth, and then show that the result can be extended by continuity to
arbitrary spectral curves.

\subsection{A model for the spectral curve}

We begin with a lemma giving a standard form for smooth 
spectral curves of hyperbolic 2-monopoles.

\begin{lemma}
    If the spectral curve of a 2-monopole is smooth, its ${\rm PSL}_2
    \mathbb{C}$-orbit contains a curve $\Sigma \subset Z$ of the form
    \begin{equation}\label{standardsc}
	\psi(w,z)=w^2 z^2 +\frac{u^2-2uv+4}{2(u-v)} w z
	-\frac{u^2-4}{4(u-v)}(w^2 + z^2)+1=0.
    \end{equation}
    Here, $u,v \in \;]2,\infty[$ depend only on the monopole mass and
    on an internal modulus, and they satisfy
    \begin{equation} \label{uvpositivity}
	u^2-2uv+4>0.
    \end{equation}
\end{lemma}
\begin{proof}
    The spectral curve $\Sigma$ of a 2-monopole is
    the vanishing locus of a polynomial of the form
    \begin{eqnarray}\label{Pkeq2}
	\psi(w,z)&=&\langle v_0,v_2\rangle w^2 z^2 - \sqrt{2}\langle
	v_1,v_2\rangle (w^2 z+ w z^2)+
	||v_0||^2(w^2+z^2)- \nonumber \\
	&& -2||v_{1}||^2 wz +\sqrt{2}\langle
	v_2,v_1\rangle(w+z)+\langle v_2,v_0\rangle,
    \end{eqnarray}
    where $v_j \in \mathbb{C}^{3}$.  The intersection of $\Sigma$ with
    the diagonal $\mathbb{P}^{1}_{\Delta}\subset\mathbb{P}^{1}\times
    \mathbb{P}^{1}$, which we parametrise using the coordinate $z$, is
    then given by the zeroes of the quartic polynomial $\psi(z,z)$.  From
    (\ref{Pkeq2}) we find
    \[
    \psi(\hat{z},\hat{z})=\bar{z}^{-4}\overline{\psi(z,z)},
    \]
    which impies that these zeroes occur in antipodal pairs. 
    Moreover, these points must be distinct for $\Sigma$ to be smooth. 
    So we can use an element of ${\rm Stab}_{(0,0,1)} {\rm
    PSL}_{2}\mathbb{C}\cong{\rm SO}(3)$ to rotate them to the
    positions $\pm\sqrt{\lambda}$ and $\pm\frac{1}{\sqrt{\lambda}}$ on
    the real axis of $\mathbb{P}^{1}_{\Delta}$.  The value of
    $\lambda\in \;]0,1[$ is uniquely determined by this process, once
    the mass of the monopole has been fixed.  Thus we have shown 
    that a rotation can be applied to have
    \begin{equation}\label{simplePdiag}
	\psi(z,z)=z^4-\left(\lambda+\frac{1}{\lambda}\right)z^2 +1;
    \end{equation}
    in this step, we have used the freedom of multiplying $\psi$ by an
    overall factor in $\mathbb{C}^{*}$.  The most general form for
    $\psi(w,z)$ consistent with (\ref{simplePdiag}) is
    \[
    \psi(w,z)=w^2 z^2+\Lambda^{2} wz-
    \frac{1}{2}(\lambda+\frac{1}{\lambda}+\Lambda^2) (w^2 + z^2)+1=0.
    \]
    Invariance under the reality map only constrains $\Lambda^2$ to be
    a real number.  However, given this, the positivity condition on
    spectral curves can easily be seen to be equivalent to $\Lambda^2
    >0$.  To complete the proof, we need only to observe that (by an
    easy continuity argument) the map $(u,v) \mapsto
    (\lambda+\frac{1}{\lambda},\Lambda^2)$ defined by
    \begin{equation}\label{bijection}
	\begin{array}{rcl}
	    \left\{ (u,v)\in \;]2,\infty[^2:u^2-2uv+4>0 \right\}&
	    \longrightarrow &
	    ]2,\infty[\times ]0,\infty[\\
	    (u,v) & \longmapsto & \left(\frac{uv-4}{u-v},
	    \frac{u^2-2uv+4}{2(u-v)}\right)
	\end{array}
    \end{equation}
    is bijective, and that
    \begin{equation}\label{ugrv}
	u>v
    \end{equation}
    is always satisfied in the domain of (\ref{bijection}).
\end{proof}

Our task is to calculate the dependence of $u$ and $v$ on the mass $m$
and an extra real parameter.  As a preliminary step, we need to obtain
a good description of the family of elliptic curves $\Sigma\subset
\mathbb{P}^1\times\mathbb{P}^1$ given by (\ref{standardsc}).  We
introduce the map
\begin{equation} \label{projpi}
    \begin{array}{rcl}
	\pi: \mathbb{P}^1 \times \mathbb{P}^1 & \longrightarrow &
	\mathbb{P}^2 \\
	\left([w_0:w_1],[z_0:z_1]\right)& \longmapsto & [w_0 z_0:w_0 z_1 +w_1
	z_0:w_1 z_1].
    \end{array}
\end{equation}
which can be regarded as a projection onto the space of orbits of the 
automorphism
\begin{equation} \label{sigma+}
    \sigma_{+}: (w,z) \mapsto (z,w).
\end{equation}
This map has order two, commutes with the reality structure $\sigma$ 
defined in (\ref{reality}), and can be written as
\[
\sigma_{+}=\sigma \circ \sigma_{-}= \sigma_{-} \circ \sigma,
\]
where 
\begin{equation}
    \sigma_{-}:(w,z) \mapsto (\hat{w},\hat{z});
\end{equation}
$\sigma_-$ is induced on $\bpxp$ by the parity
transformation on hyperbolic space
\begin{equation} \label{parity}
    {\cal P}_{(0,0,1)}:(x_1,x_2,x_3)\mapsto
    \left(-\frac{x_1}{x_1^2+x_2^2},\frac{x_2}{x_1^2+x_2^2},\frac{1}{x_3}\right),
\end{equation}
which can be described as reflection across $(0,0,1)$.  Since ${\cal
P}_{(0,0,1)}$ reverses the orientation of $H^3$, $\sigma_{-}$ is
antiholomorphic, and it is a reality structure on $\bpxp
$ or $Z$ alternative to (\ref{reality}).  Thus there is a Vierergruppe
of diffeomorphisms of twistor space associated with $(0,0,1)$ (and
likewise with any other point of $H^3$),
\[
{\cal V}_{(0,0,1)}:= \{{\rm id}, \sigma, \sigma_-, \sigma_+ \} \cong
\bz_2 \oplus \bz_2,
\]
which is naturally $\bz_2$-graded by (anti)holomorphicity.  That
$\sigma_{+}$ should arise as a symmetry of spectral curves
(\ref{Pkeq2}) of centred 2-monopoles is unsurprising, as we expect
their fields to be symmetric under the parity transformation
(\ref{parity}).

The fixed points of $\sigma_{+}$ form the diagonal
$\mathbb{P}^1_{\Delta}$ of $\mathbb{P}^1 \times \mathbb{P}^1$ (i.e.\
the star at $(0,0,1)$), and $\sigma_{+}^2={\rm id}$.  Thus $\pi$ is
two-to-one on $\mathbb{P}^1 \times \mathbb{P}^1 -
\mathbb{P}^1_{\Delta}$ and one-to-one on $\mathbb{P}^{1}_{\Delta}$. 
In homogeneous coordinates ${Z_0}$, ${Z_1}$ and ${Z_2}$ for
$\mathbb{P}^2$, $\pi(\mathbb{P}^1_\Delta)$ is the conic given by
\begin{equation}
    C(Z_0,Z_1,Z_2):=Z_1^2-4Z_0 Z_2=0.
\end{equation}
The image of $\Sigma$ under $\pi$ is given by the equation
\[
\tilde{\psi}(\zeta_1,\zeta_2):= \zeta_2^2-\frac{u^2-4}{4(u-v)}\zeta_1^2 
+ u \zeta_2+1=0
\]
in affine coordinates $\zeta_1=\frac{Z_1}{Z_0}=w+z$ and
$\zeta_2=\frac{Z_2}{Z_0}=wz$.  This is also a plane conic, so we can
find a rational parametrisation $f:\mathbb{P}^1 \rightarrow
\pi(\Sigma)$ for it.  In fact, if we write
\begin{equation}\label{paraf}
    f(t)=[1-\alpha t^2:2\alpha\beta t:t^2-\alpha],
\end{equation}
we will have
\[
f(\mathbb{P}^{1})=\pi(\Sigma)
\]
if and only if we choose $\alpha$ and $\beta$ such that the conditions
\begin{equation}\label{alpha}
    u=\alpha+\frac{1}{\alpha}
\end{equation}
and
\begin{equation}\label{beta}
    u-v = \alpha \beta^2
\end{equation}
are satisfied; $\alpha$ and $\beta$ are then real.  Notice that the
equation (\ref{alpha}) determining $\alpha$ indeed gives two real
solutions of the form $\alpha, \frac{1}{\alpha}$ since $u>2$, and
because of the invariance under $\alpha \mapsto \frac{1}{\alpha}$ we
are free to assume that
\[
\alpha>1,
\]
which we shall do from now on; that $\beta$ is real then follows from
(\ref{ugrv}).  We now realise $\Sigma$ as a double cover of the
$\mathbb{P}^1$ where $f$ is defined, branching on the set of four
points $f^{-1}(\pi(\Sigma)\cap\pi(\mathbb{P}^1_{\Delta}))$.  One way
to do this is to regard $\Sigma$ as the Riemann surface of the global
function of $t$ extending the germ
\begin{equation} \label{germ}
    t\mapsto \sqrt{F(t)}
\end{equation}
at $t=0$, where
\begin{equation}\label{defF}
    F(t):=C\circ f (t)=4 \alpha\left( t^4-v t^2+1\right).
\end{equation}
(In (\ref{germ}) and henceforth, we use $\sqrt{\cdot}$ to denote the
principal branch of the square root.)  The branch points of
$\Sigma\rightarrow \mathbb{P}^1$ are the zeroes of $F$, and they are
of the form $\pm\sqrt{\kappa}$, $\pm \frac{1}{\sqrt{\kappa}}$ with
\begin{equation}\label{kappa}
    \kappa+\frac{1}{\kappa}=v.
\end{equation}
Again, $v>2$ ensures that $\kappa$ is real and positive, and by
invariance of (\ref{kappa}) under $\kappa \mapsto \frac{1}{\kappa}$ we
may assume from now on that
\begin{equation}\label{kappalo1}
    0<\kappa<1.
\end{equation}

The elliptic curve $\Sigma$ given by (\ref{standardsc}) can now be
realised as a two-sheeted cover of the projective line parametrised by
$t$ as follows (cf.~Figure~1).  Introduce branch cuts on the segments
$[-\frac{1}{\sqrt{\kappa}},-\sqrt{\kappa}]$ and
$[\kappa,\frac{1}{\sqrt{\kappa}}]$, and label the two sheets as (i)
and (ii), where sheet (i) is the one on which the germ (\ref{germ})
(which takes a positive value at $t=0$) is defined.  Clearly, on sheet
(i)/(ii), the analytic continuation of (\ref{germ}) then takes
negative/positive values on $]-\infty,-\frac{1}{\sqrt{\kappa}}[$ and
$]\frac{1}{\sqrt{\kappa}},+\infty[$, and positive/negative values on
$]-\sqrt{\kappa},{\sqrt{\kappa}}[$, respectively.

\begin{figure}[ht] 
    \begin{center}
	\vspace{.8cm}
	\input{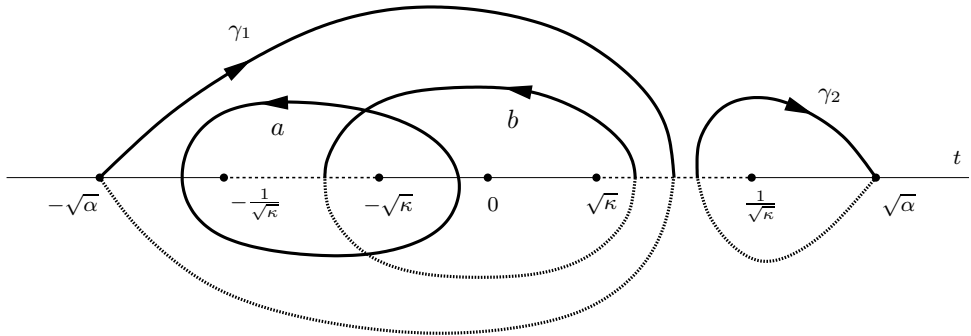}
	\small{
	\caption{Hyperbolic 2-monopole: contours of integration in a 
             realisation of $\Sigma$}
	}
    \end{center}
\end{figure}

To complete the correspondence between our descriptions of $\Sigma$ as
the Riemann surface of the germ (\ref{germ}) and as a hypersurface in
$\mathbb{P}^{1}\times\mathbb{P}^{1}$, we just need to make the
identification of the two descriptions at a single point.  We choose
this point $p_1$ to be an element of the fibre of $\Sigma \rightarrow
\mathbb{P}^1$ over $t=-\sqrt{\alpha}$.  In the description of $\Sigma$
by analytic continuation, we specify $p_1$ by saying that it lies on
sheet (i).  In the description $\Sigma \subset \mathbb{P}^{1} \times
\mathbb{P}^{1}$, we will need to specify the coordinates $(w,z)$ of
$p_1$.  Notice that from (\ref{paraf}) we have that when
$t=\mp\sqrt{\alpha}$
\[
wz=\frac{(\mp\sqrt{\alpha})^2-\alpha}{1-\alpha(\mp\sqrt{\alpha})^2}=0,
\]
so either $w=0$ or $z=0$; we specify $p_1$ by saying that it has
$z=0$.  Then we find from (\ref{standardsc}) that $w=\pm
2\sqrt{\frac{u-v}{u^2-4}}$; we must set
\begin{equation}\label{p1}
    p_{1}:=\left( 2 \sqrt{\frac{u-v}{u^2-4}} ,0 \right),
\end{equation}
since (\ref{projpi}) implies that the point
\begin{equation}\label{p2}
    p_2 := \left( -2 \sqrt{\frac{u-v}{u^2-4}} ,0 \right)
\end{equation}
projects to $t=+\sqrt{\alpha}$ under (\ref{projpi}).  For further
reference, we note that the other points on the fibres over
$t=\mp\sqrt{\alpha}$ have $w=0$ and $z=\pm 2
\sqrt{\frac{u-v}{u^2-4}}$, and we label them as
\begin{equation}\label{q1q2}
    q_1:=\left( 0,2\sqrt{\frac{u-v}{u^2-4}}\right), \quad
    q_2:=\left( 0,-2\sqrt{\frac{u-v}{u^2-4}}\right).
\end{equation}
By construction, $p_1$ is on sheet (i), so $q_{1}$ must be on sheet
(ii) as it belongs to the same fibre.  Moreover, one can write
\[
w(t)=\frac{Z_1(t) \pm \sqrt{F(t)}}{2Z_0 (t)}
\]
where the signs $\pm$ distinguish between the two sheets; this
equation implies that, on a given sheet, if $w=0$ at
$t=\mp\sqrt{\alpha}$, then $w\ne 0$ at $t= \pm \sqrt{\alpha}$, and the
same is true for $z$.  So we conclude that $p_2$ is on sheet (ii),
while $q_2$ is on sheet (i).

We want to discuss the conditions characterising a spectral curve
$\Sigma$ from the point of view of the function theory on $\Sigma$,
and for that we shall need to fix generators for the spaces of 1-forms
and 1-cycles on the elliptic curve.  A global holomorphic 1-form on
$\Sigma$ is given by the standard formula
\begin{equation}\label{omega}
    \omega=\frac{{\rm d}t}{\pm\sqrt{F(t)}}
\end{equation}
which follows from taking Poincar\'e residues of
(\ref{standardsc})~\cite{GriHar}; again, the signs label sheets, and
we make the convention of taking the top sign on sheet (i) near $t=0$. 
To describe a standard basis $(a,b)$ for
$H_1(\Sigma;\mathbb{Z})\cong\mathbb{Z}^{2}$ we draw representatives
for $a$ and $b$ in Figure~1.  The convention is that paths on sheet
(i) are drawn with continuous lines, whereas paths on sheet (ii) are
drawn with dotted lines.

\subsection{Reciprocity on $\Sigma$} \label{sec:reciproc}

To make the connection between the algebraic-geometric condition 2.\
and complex analysis on $\Sigma$, we shall use the reciprocity law for
differentials of first and third kinds on a compact Riemann surface
(cf.~\cite{GriHar}, pp.~229--230).  This technique is in the same
spirit of previous work on euclidean monopoles, where a different
reciprocity law has been applied to deduce constraints on the
coefficients of polynomials defining spectral
curves~\cite{Hur,ErcSinMBF,HouManRom}.

We start by rephrasing $\left.  L^{2m+k}\right|_{\Sigma}\cong {\cal
O}_{\Sigma}$ in terms of the existence of a global holomorphic
trivialisation for the line bundle $L^{2(m+1)}$ on $\Sigma$, or
equivalently a nowhere-vanishing holomorphic section $\xi \in
H^{0}(\Sigma,L^{2(m+1)})$.  Such a section is locally
represented by nowhere-vanishing holomorphic functions $\xi_j$ defined
on the basic open sets $U_j \cap \Sigma$ ($j=1,2$) (cf.\
(\ref{U1U2})), and which are patched together using the appropriate
transition function (\ref{trans}):
\begin{equation}\label{trivializat}
  \xi_{1}(w,z)=  \left( \frac{w}{z} \right)^{2(m+1)}  \xi_{2}(w^{-1},z^{-1}).
\end{equation}
To begin with, this patching condition should hold on $U_{1}\cap U_{2}
\cap \Sigma$, which consists of $\Sigma$ with eight points deleted,
namely the points of the form $(w,0)$, $(w,\infty)$, $(0,z)$ and
$(\infty,z)$.  There are two of each, since the polynomial $\psi(w,z)$
defining $\Sigma$ has bidegree $(2,2)$; for example, the points of the
form $(0,z)$ are $p_1$ and $p_2$ given by equations (\ref{p1}) and
(\ref{p2}).  Taking logarithmic differentials in both sides of
(\ref{trivializat}), we obtain the relation
\begin{equation}\label{concretely}
    \omega_{1}=2(m+1)\left( \frac{{\rm d}w}{w} - \frac{{\rm d}z}{z}\right) +
    \omega_{2}
\end{equation}
where
\begin{equation}\label{omegaj}
    \omega_{j}:={\rm d} \log \xi_{j}=\frac{{\rm d} \xi_j}{\xi_j}\qquad j=1,2.
\end{equation}
The definition (\ref{omegaj}) implies the following integrality
property on periods:
\begin{equation}\label{intperiods}
    \oint_{a}\omega_j , \oint_{b}\omega_j \; \in\;  2 \pi i\; \mathbb{Z}. 
\end{equation}
We set
\begin{equation} \label{ells}
    \ell_1:=\frac{1}{2\pi i}\oint_b \omega_1,\qquad
    \ell_2:=-\frac{1}{2 \pi i}\oint_a \omega_1.
\end{equation}
Notice that we can replace $2(m+1)$ by $s$ and go through the same
argument to conclude that condition 3.\ can be rephrased as:
\begin{equation}\label{primitive}
    (\ell_1,\ell_2)\quad \mbox{must be primitive in} \; \mathbb{Z}^{2}. 
\end{equation}

Each $\omega_j$ is a holomorphic 1-form on $U_j\cap \Sigma$ (since
$\xi_j$ are holomorphic and never vanish there), but they extend to
global meromorphic 1-forms on $\Sigma$ as a consequence of
(\ref{concretely}).  To see this, notice that $\omega_2$ is
holomorphic on a small neighbourhood of
\[
\Sigma-U_1=\left\{ p_1,p_2,q_1,q_2\right\}
\]
in $\Sigma$, and then (\ref{concretely}) implies that $\omega_1$ must
have the same principal part as $2(m+1)\left(
\frac{{\rm d}w}{w}-\frac{{\rm d}z}{z}\right)$ on this set.  Since $w$ is a good
coordinate on $\Sigma$ around $p_1$ and $p_2$, while $z$ is a good
coordinate around $q_1$ and $q_2$, we can conclude that $\omega_1$ has
simple poles at each of these points, and we find
\begin{equation}\label{residues}
    {\rm Res}_{p_j}(\omega_1)=-2(m+1) \;\quad \mbox{and} \;\quad {\rm
    Res}_{q_j}(\omega_1)=2(m+1), \;\quad j=1,2.
\end{equation}
A similar analysis can be done for $\omega_2$.

In classical terminology, a holomorphic 1-form (such as $\omega$) is
called a differential of the first kind, while a meromorphic 1-form
with only single poles (such as $\omega_1$) is called a differential
of the third kind.  A standard result of complex analysis on compact
Riemann surfaces is the reciprocity law~\cite{GriHar}
\begin{equation} \label{reciprocity}
    \left|\begin{array}{cc}
    \oint_a \omega &  \oint_a\omega_1 \\
    \oint_b \omega & \oint_b\omega_1 \end{array}\right| = 2 \pi i
    \sum_{q} {\rm Res}_{q}(\omega_1) \int_{p_0}^{q}\omega.
\end{equation}
Here, $p_0\in \Sigma$ is any basepoint, the integration paths on the
right-hand side can be deformed to avoid the paths representing the
homology basis, and the sum is over the set of poles of $\omega_1$.
Equation (\ref{reciprocity}) relates two elements of the covering space 
$H^{1}(\Sigma,{\cal O}_\Sigma) \cong T_{{\cal O}_\Sigma}{\rm Jac}(\Sigma)$
to the jacobian variety of $\Sigma$.
On a complex curve of higher genus $g$, the left-hand-side of the reciprocity
relation would have a
sum over conjugate pairs of a standard (symplectic) basis of
$H_1(\Sigma;\mathbb{Z})$, and there would be $g$ equations. Note that by Liouville's
theorem
\[
\sum_{q}{\rm Res}_{q}(\omega_1)=0,
\]
which explains why the right-hand side of (\ref{reciprocity}) is in 
fact independent of $p_0$.

In our context, using (\ref{ells}) and (\ref{residues}), equation
(\ref{reciprocity}) can be written as
\begin{equation}\label{recip}
    \ell_1\oint_a \omega + \ell_2\oint_b \omega =2(m+1)
    \left( \int_{p_1}^{q_1} \omega + \int_{p_2}^{q_2}\omega \right).
\end{equation}
We note in passing that elements of 
$H^{0}(\Sigma,K_\Sigma)^{*} \cong H^1(\Sigma,{\cal O}_{\Sigma})$ are
realised in different ways in the two sides of this equation, namely as
a pairing (via $\oint$) with a 1-cycle 
$\ell_1 a + \ell_2 b \in H_{1}(\Sigma;\bz)$ in the left-hand side, and
as a multiple of an Abel sum associated to the divisor 
$q_1 + q_2 -p_1 - p_2 \in {\rm Div}_0(\Sigma)$ representing the class
${\cal O}(1,-1)$ in the right-hand side;
in fact, if we interpret the integrals in terms of the Abel--Jacobi map, 
and use congruence on the period lattice of $\Sigma$,
we can read (\ref{recip}) as a rather transparent statement of
$L^{2(m+1)}|_\Sigma$ being trivial.  
Of course, there is a freedom of changing the subscripts of the $p_j$
and $q_j$ in this expression, but we find it convenient to choose
paths of integration as in (\ref{recip}).  More precisely, we choose
paths $\gamma_j$ ($j=1,2$) on $\Sigma$ starting at $p_j$ and ending at
$q_j$ as illustrated in Figure~1, but defined only as a class in the relative
homology of $(\Sigma,\{p_j,q_j\})$.  Notice that the locations of the
projections $t=\pm\sqrt{\alpha}$ of the poles of $\omega_1$ with
respect to the branch points are as shown, since
\[
u>v \Leftrightarrow \alpha+\frac{1}{\alpha} > \kappa + 
\frac{1}{\kappa} \Rightarrow \alpha > \frac{1}{\kappa}.
\]

To understand what constraint (\ref{recip}) imposes on the
coefficients of $\Sigma$, we have to compute the integrals in this
equation.  Under our conventions, it is clear that all of them are
real except for the period $\oint_{a}\omega$, which is pure imaginary. 
Therefore $\ell_1 =0$, and (\ref{primitive}) imposes $\ell_2=\pm 1$. 
So we only need to evaluate
\[
\oint_b \omega = -2 I_1,
\]
\[
 \int_{\gamma_1}\omega=2(I_1-I_2)
\]
and
\[
\quad \int_{\gamma_2}\omega=-2I_2,
\]
where
\[
I_1:=\int_{-\sqrt{\kappa}}^{\sqrt{\kappa}}\frac{{\rm d}t}{\sqrt{F(t)}},
\qquad
I_2:=\int_{\frac{1}{\sqrt{\kappa}}}^{\sqrt{\alpha}}\frac{{\rm d}t}{\sqrt{F(t)}}.
\]
In terms of standard notation for elliptic integrals of the first kind
\begin{eqnarray*}
    &\D
    F(\varphi,\kappa):=\int_{0}^{\varphi}\frac{{\rm d}\theta}{\sqrt{1-\kappa^2
    \sin^2 \theta}},&\\
    &\D K(\kappa):=F\left(\frac{\pi}{2},\kappa\right),&
\end{eqnarray*}
we calculate
\[
I_1=\sqrt{\frac{\kappa}{\alpha}}K(\kappa)
\]
and
\[
I_2= \frac{1}{2}\sqrt{\frac{\kappa}{\alpha}}\left( K(\kappa) -
F\left({\rm arcsin\frac{1}{\sqrt{\alpha \kappa}}},\kappa\right)
\right).
\]
Substituting in equation (\ref{recip}), we find that $\ell_1=-1$ and
\begin{equation}\label{massrel}
    F\left( {\rm arcsin} \frac{1}{\sqrt{\alpha \kappa}},\kappa
    \right)= \frac{K(\kappa)}{2(m+1)}.
\end{equation}

\subsection{Mass parametrisation}

Equation (\ref{massrel}) can be solved for $\alpha$ to give
\begin{equation}\label{alphaofkm}
    \alpha=\frac{1}{\kappa \;{\rm sn}^{2}\left(
    \frac{K(\kappa)}{2(m+1)},\kappa\right)}
\end{equation}
in terms of Jacobi's sine amplitude function.  (In the following, we
will often drop the modulus argument $\kappa$ in jacobian elliptic
functions.)  Using (\ref{kappa}), (\ref{alphaofkm}) and standard
algebra of jacobian functions~\cite{ByrFri}, we find
\begin{equation}\label{1stcoeff}
    \frac{u^2-2uv+4}{2(u-v)}=\frac{2}{\kappa}\; {\rm cs}\left(
    \frac{K(\kappa)}{m+1}\right) {\rm ds}\left( \frac{K(\kappa)}{m+1}
    \right)
\end{equation}
and
\begin{equation}\label{2ndcoeff}
    \frac{u^2-4}{4(u-v)}=\frac{1}{\kappa}\;{\rm
    ns}^2\left(\frac{K(\kappa)}{m+1} \right).
\end{equation}
So we have shown that equation (\ref{standardsc}) defines a 2-monopole
of mass $m$ if and only if (\ref{1stcoeff}) and (\ref{2ndcoeff}) hold. 
Moreover, we shall see in section~\ref{secaxialsym} below that we can
extend our results by continuity to the $\kappa \rightarrow 0$ limit,
where the spectral curve becomes reducible and singular.  In other
words, we have the following result.

\begin{proposition}\label{propstform}
    A spectral curve of a 2-monopole of mass $m$ centred at $(0,0,1)
    \in H^3$ can be rotated to the form
    \begin{equation} \label{massparam}
	\kappa \;{\rm sn}^2 \rho \; (w^2 z^2 +1) + 2 \; {\rm cn}\, \rho \;
	{\rm dn} \, \rho\;  wz -(w^2 + z^2) = 0,
    \end{equation}
    where $0\ \le \kappa<1$ and
    \begin{equation} \label{upsilon}
	\rho=\frac{K(\kappa)}{m+1}.
    \end{equation}
\end{proposition}

When $m$ is a rational number, the coefficients of (\ref{massparam})
turn out to be algebraic over $\mathbb{Q}(\kappa)$. For example,
if $m=1$ we can use bisection formulae for jacobian elliptic
functions~\cite{ByrFri} to write the spectral curve in the form
\[
\kappa(w^2 z^2 + 1)+2\kappa' \sqrt{1+\kappa'} wz-(1+\kappa')(w^2+z^2)=0,
\]
where $\kappa':=\sqrt{1-\kappa^2}$.
We shall encounter later a similar phenomenon occurs in examples of spectral curves of higher charge.

\section{Limiting cases of 2-monopoles}\label{seclimcases}

In this section, we examine and interpret the limits of our result in
Proposition~\ref{propstform} as $\kappa \rightarrow 0$, $\kappa
\rightarrow 1$, $m \rightarrow 0$ and $m \rightarrow \infty$.

\subsection{Axial symmetry}\label{secaxialsym}

When we let $\kappa \rightarrow 0$,
\begin{eqnarray*}
    &{\rm sn}\, \rho \rightarrow \sin \rho, &\\
    &{\rm cn}\, \rho \rightarrow \cos \rho, & \\
    &{\rm dn}\, \rho \rightarrow 1.& 
\end{eqnarray*}
whereas (\ref{upsilon}) implies
\[
\rho \rightarrow \frac{\pi}{2(m+1)}.
\]
Thus (\ref{massparam}) becomes
\begin{equation}\label{axsim}
    w^2  -2 \cos \left(\frac{\pi}{2(m+1)} \right) w z + z^2 =0.
\end{equation}
This is the spectral curve of an axially symmetric 2-monopole at
$(0,0,1)$, as computed in reference~\cite{MurNorSinHS}.  This curve is
reducible, since the polynomial in (\ref{axsim}) factorises as
\[
\left(w-e^{\frac{i \pi}{2(m+1)}}z \right)
\left(w-e^{-\frac{i \pi}{2(m+1)}}z \right).
\]
The reduced components are two projective lines in
$\mathbb{P}^1\times\mathbb{P}^1$, intersecting at the points $(0,0)$
and $(\infty,\infty)$ in the diagonal, which are related by the real
structure $\sigma$.  Notice that the lines
\[
w= e^{\pm\frac{i \pi}{2(m+1)}} \, z
\]
are real and can be thought of as stars at the complex conjugate
points $(0,0,e^{\pm\frac{i \pi}{4(m+1)}})$ (cf.\ equation
(\ref{star0}) for a star at $(0,0,x_3)\in H^3$).  This situation is
analogous to axially symmetric monopoles in euclidean
space~\cite{HitManMur}.

We conclude that we can extend the range of $\kappa$ from $0<\kappa<1$
to $0\le \kappa <1$, as in Proposition~\ref{propstform}.

\subsection{Large separation}

To study the limit $\kappa \rightarrow 1$, we use that~\cite{ByrFri}
\begin{eqnarray*}
    & {\rm sn} \, \rho \rightarrow {\rm tanh} \, \rho, & \\
    & {\rm cn} \, \rho \rightarrow  {\rm sech}\, \rho, & \\
    & {\rm dn} \, \rho \rightarrow {\rm sech}\, \rho, & 
\end{eqnarray*}
together with
\[
\rho \rightarrow + \infty
\]
from equation (\ref{upsilon}).  We find that the spectral curve
(\ref{massparam}) then degenerates to
\begin{equation}\label{infsep}
    (w^2-1)(z^2-1) = 0.
\end{equation}
This reduces to four real lines in $\mathbb{P}^1 \times \mathbb{P}^1$. 
The two lines $w=\mp 1$ and $z=\pm 1$ together can be regarded as a
limiting star at the point $(\pm 1,0,0) \in \partial H^3$, for either
choice of signs; these two lines intersect at the point $(\mp 1, \pm
1)\in \mathbb{P}^{1}_{\bar\Delta}$, respectively.  We interpret
(\ref{infsep}) as the spectral curve of two 1-monopoles that are
infinitely separated at each of the ends of the geodesic $\mu
\circ\nu^{-1} (1,1)=\mu\circ\nu^{-1}(-1,-1)$ of $H^3$.

The analysis of the limiting examples $\kappa \rightarrow 0$ and
$\kappa \rightarrow 1$ suggests that $\kappa$ can roughly be thought
of as a parameter of the separation of two single monopole cores in
configurations of 2-monopoles centred at $(0,0,1)$.  As $\kappa$
varies from $1$ to $0$, the cores approach symmetrically along the
geodesic given by $x_2=0$, $x_1^2 + x_3^2=1$ from infinite distance to
coincidence at $(0,0,1)$ (the axially symmetric configuration). 
Strictly speaking, the description of a 2-monopole configuration in
terms of two superposed 1-monopoles is only appropriate in the
asymptotic limit of large separation, and becomes worse and worse as
$\kappa$ decreases.  This is why the 1-monopole cores in the axially
symmetric configuration are found to be located at points of the
complexification of $H^3$ rather than at the centre $(0,0,1)$ itself.

\subsection{2-Nullarons}\label{sec:2nullar}

If we let $m\rightarrow 0$, then~\nocite{ByrFri}
\begin{eqnarray*}
    & {\rm sn}\,\rho \rightarrow {\rm sn}\, K(\kappa) = 1,& \\
    & {\rm cn}\,\rho \rightarrow {\rm cn}\, K(\kappa) = 0,& \\
    & {\rm dn}\, \rho \rightarrow {\rm dn}\, K(\kappa) =
    \sqrt{1-\kappa^2},&
\end{eqnarray*}
and the spectral curve (\ref{massparam}) becomes
\begin{equation}\label{2nullaron}
    \kappa(w^2 z^2+1)-(w^2 + z^2)=0. 
\end{equation}
This result can be obtained by more direct means.  In fact, (\ref{2nullaron})
is the standard $\mathbb{Z}_{2}$-symmetric curve encoding a solution
of the Potts model~\cite{AtiMMYB}, and can be computed as
\begin{equation}\label{nullaroncurve}
    \widehat{R(w)}=R(\hat{z})
\end{equation}
with
\[
R(z)=\frac{\sqrt{1-\kappa^2}}{z^2-\kappa}.
\]

More generally, equation (\ref{nullaroncurve}) produces all the
spectral curves of massless monopoles (nullarons) of charge $k$ from
rational maps $R:\mathbb{P}^1 \rightarrow \mathbb{P}^1$ of degree $k$.

\subsection{Euclidean 2-monopoles} \label{sec:euclim2}

In the limit $m\to\infty$,  which is equivalent to $\rho\to 0$, the curve (\ref{massparam}) tends to 
two copies of the diagonal $\bp^1_\Delta \subset \bp^1 \times \bp^1$,
with equation $(w-z)^2=0$. However, at the next asymptotic order, we can
recover the spectral curve of a euclidean monopole embedded in 
$T\bp^1_\Delta$.

To explain this, we must first make more precise what is meant by the euclidean limit of a hyperbolic monopole.  As discussed in the introduction, rescaling the hyperbolic metric to have larger and larger curvature radius
$R$ is equivalent to rescaling the fields (equivalently, the spectral curves), and in particular their mass, while keeping the metric constant; the infinite radius limit $R \rightarrow \infty$ can then be thought of as an infinite mass 
limit $m \rightarrow \infty$, to be interpreted as a euclidean monopole.  
Implicit in this rescaling is the choice
of a point in $H^3$.  We choose to rescale around the centre $(0,0,1)\in H^3$, and this forces us to consider only centred hyperbolic monopoles with limit centred euclidean monopoles.
More generally, rescaling around a different point in $H^3$ would lead
to limit spectral curves concentrated around the corresponding star in
$\bpxp$.

Recall that a point $(w,z)\in\bpxp$ is interpreted as a geodesic in $H^3$ running from $\hat{w}$ to $z$, both points on the sphere at infinity. From the perspective of the centre $(0,0,1)$, the geodesic is viewed as the vector
pointing to its closest point from $(0,0,1)$ and the (orthogonal) tangent 
direction at that point. Now take a 
sequence of geodesics in $H^3$ in the spectral curves converging to a geodesic through $(0,0,1)$, or equivalently a sequence of points $(w_m,z_m)\in\bpxp$ for $m\to\infty$ with limit $(w_{\infty},z_{\infty})=(z_{\infty},z_{\infty})$ on the diagonal.  
For each value of $m$, rescale the radius of curvature of the metric
on $H^3$ by $m$. Then a simple geometric argument (involving similarity of infinitesimal triangles on the 2-plane containing the sequence of geodesics) shows that, in the limit, one obtains the euclidean geodesic
\begin{equation}  \label{eq:geodlim}
(w_m,z_m)\to(\eta,\zeta)=\left( \lim_{m\to\infty} m(z_m-w_m), 
z_{\infty}\right),
\end{equation}
where we are using the standard coordinates 
on $T\bp^1_{\Delta}$ introduced by 
Hitchin~\cite{HitMG}.
Given the $\rho \rightarrow 0$ asymptotics
\[
{\rm sn}^2 (\rho,\kappa) = \rho^2 + O(\rho^3)
\] 
and
\[
{\rm cn}(\rho ,\kappa)\, {\rm dn}(\rho,\kappa) =
1-\left( \frac{1+\kappa^2}{2}\right)\rho^2 
+ O(\rho^3),
\]
we can write
\begin{eqnarray*}
\eta^2& =&\lim_{m\to\infty}m^2(w_m-z_m)^2\\
&=&K(\kappa)^2\lim_{\rho\to 0}
\frac{(w_m-z_m)^2}{\rho^2}\\&=&
K(\kappa)^2\lim_{\rho\to 0}\frac{(2 \; {\rm cn}\, \rho \;
{\rm dn} \, \rho\; -2)w_m z_m +\kappa \;{\rm sn}^2 \rho \; (w_m^2 z_m^2 +1)}{\rho^2}\\&=&
K(\kappa)^2\lim_{m\to \infty}\left(-(1+\kappa^2)w_m z_m+\kappa(w_m^2 z_m^2+1)\right)\\&=&
K(\kappa)^2(-(1-\kappa^2)\zeta^2+\kappa(\zeta^4-1)),
\end{eqnarray*}
where we have used (\ref{eq:geodlim}) in the first and last steps.  Hence
we obtain the curve in $|{\cal O}_{T\bp^1}(4)|$ 
\begin{equation}\label{eucllimk2}
\eta^2-K(\kappa)^2(\zeta^2-\kappa)(\kappa\zeta^2-1)=0
\end{equation}
as the euclidean limit of (\ref{massparam}). 
To compare this with the spectral curve of a generic euclidean 2-monopole~\cite{Hur},
as given in reference~\cite{ManSut} (say),
\begin{equation}\label{standk2}
\eta^2-\frac{K(k)^2}{4}\left(k^2(\zeta^4+1)-2(2-k^2)\zeta^2 \right)=0,
\end{equation}
where $0< k<0$, one can start by relating $\kappa$ and $k$ by computing the
corrdinate $\zeta$ at the four intersection points of
(\ref{eucllimk2}) and (\ref{standk2}) with the zero section $\eta=0$ of
$T\bp^1$ (i.e.\ the branch points of $\Sigma \rightarrow \bp^1$).  In (\ref{eucllimk2}) we find $\zeta=\pm\sqrt{\kappa}, 
\pm\frac{1}{\sqrt{\kappa}}$, whereas for (\ref{standk2}) the solutions have
the same form but are parametrised differently, namely
\[
\kappa=\frac{2}{k^2}-1-\sqrt{\left(\frac{2}{k^2}-1 \right)^2-1} \quad
\Rightarrow \quad
k=\frac{2 \sqrt{\kappa}}{1+\kappa}.
\]
Using a descending Landen's transformation~\cite{FriEF}, we can 
now write
\[
K(k)=K\left( \frac{2\sqrt{\kappa}}{1+\kappa} \right)=(1+\kappa) K(\kappa)
\]
and conclude that (\ref{eucllimk2}) and (\ref{standk2}) simply give two 
different parametrisations of the same curve. An alternative way of checking 
that our limit curve is correct is to start from our parametrisation
of the branch points and use the characterisation of spectral curves in
section~3 of~\cite{HouManRom} to deduce (\ref{eucllimk2}).

\section{Platonic monopoles in hyperbolic space} \label{sec:plato}

For charge $k>2$, the spectral curves are of higher genus $(k-1)^2>1$
and depend on $4k-3$ internal moduli, so it becomes a difficult task
to compute any one of them, let alone obtain a complete picture of the
whole moduli space.  A technique that has been fruitful is to impose
invariance of the monopoles under certain isometries~\cite{HitManMur};
in some cases, this cuts the number of moduli down to a manageable
number, while important features of the moduli space are preserved. 
For instance, such symmetry constraints define totally geodesic
subsets of the moduli space, whose geodesic flow is simply a
restriction of the ambient geodesic flow; moreover, one expects
these subsets to carry significant information about the topology of the
whole moduli space.  In the euclidean case, beautiful examples of
scattering of symmetric multi-monopoles have been found in this
way~\cite{HitManMur,ManSut}, as well as new insight into the structure
of the monopole fields themselves~\cite{SutMZ}.

Usually, the construction of spectral curves of symmetric monopoles on
euclidean space has relied on the study of Nahm's equations.  For
hyperbolic monopoles of general mass this route cannot be taken, as we
do not yet have enough understanding of the generalisation of Nahm's
equations appropriate for the problem.  In the following sections, we
shall obtain results about symmetric hyperbolic monopoles by attacking
the problem directly via the geometry of spectral curves.  
Although our methods apply more generally,
we will
restrict ourselves to monopoles with the rotational symmetry of a platonic
solid, and would like to construct associated spectral curves with arbitrary mass.
The platonic solids we
consider are the tetrahedron, the octahedron and the icosahedron, with
(special) symmetry groups isomorphic to $ {A}_4$,
$ {S}_4$ and $ {A}_5$, respectively.  Instead of the
octahedron and the icosahedron, we could have taken the dual solids
(cube and dodecahedron), which have the same group of symmetries.

We start by writing down Ans\"atze for real $(k,k)$ curves in $Z$ with
the symmetry of a given platonic solid.  Again, we may restrict
ourselves to centred curves without losing generality.  
The following elementary observation will be useful:

\begin{lemma} \label{th:fixedpts}
    Let $\Sigma \subset Z$ be a centred $(k,k)$ curve.  If $(w,z) \in
    \Sigma$ is fixed by a rotation preserving $(0,0,1)$, then $w=z$.
\end{lemma}
\begin{proof}
    The set of points of $H^3$ fixed by a (nontrivial) rotation in
    $\rm Stab_{(0,0,1)}{\rm PSL}_{2}\bc \cong {\rm SO}(3)$ is a
    geodesic through $(0,0,1)$.  Lemma~\ref{th:stars} then implies
    that the spectral lines fixed by this rotation must lie on the
    diagonal $\bp^{1}_{\Delta} \subset \bpxp$.
\end{proof}

Suppose that a platonic group $G\subset \rm Stab_{(0,0,1)}{\rm
PSL}_{2}\bc \cong {\rm SO}(3)$ has been fixed; a pictorial way of doing this is to map
the corresponding solid onto the star at $(0,0,1)$ (which we can think
of as a two-sphere centred at $(0,0,1)$ and identify with
$\bp^{1}_{\Delta}$), using central projection.
 In the following, we shall be interested in $G$-symmetric $(k,k)$ curves
$\Sigma$ for which the
space of orbits $\Sigma/G$ is an elliptic curve. Note that, if $\Sigma$ is
smooth, so too will be $E$.

\begin{proposition}\label{mincurves}
    Let $\Sigma\subset\bpxp$ be a smooth $(k,k)$ curve
    invariant under $G=A_4, S_4$ or $A_5$ symmetry with 
    quotient $\Sigma/G=E$ an elliptic
    curve.  Then $k=3$ or 4 in the $G=A_4$ case, and $k=4$ and 6
    in the $S_4$, respectively $A_5$ cases.
\end{proposition}
    \noindent {\em Remark.} In the course of the proof we will see
    that, together with the diagonal $\bp^1_\Delta\subset\bpxp$,
    the curves satisfying the conditions of the proposition are the
    smallest degree smooth $(k,k)$ curves in $\bpxp$ with
    tetrahedral, octahedral or icosahedral symmetry.
\begin{proof}    
    The quotient $\Sigma/G$ parametrises the orbits of $G$ on
    $\Sigma \subset Z$. 
    Generic orbits of $G$ on $\bp^1_\Delta$ have $|G|$ points and thus
    the same is true of generic orbits of $G$ on $Z$.  There are three
    exceptional orbits of $G$ on $\bp^1_\Delta$, and hence $Z$, given by
    orbits of vertices, (midpoints of) edges and (midpoints of) faces
    under the identification of $\bp^1_\Delta$ with the symmetric polyhedron. 
    Thus generic $G$-orbits on a $G$-symmetric $(k,k)$ curve $\Sigma$ have 
    $|G|$ points, and there are at most three exceptional orbits.  In the tetrahedral case, there are three orbits consisting of 4, 6
    and 4 points --- the orbits of vertices, edges and faces --- on the
    $(k,k)$ curve $\Sigma$.  For the octahedral and icosahedral cases,
    the exceptional orbits correspond to 6, 12 and 8 vertices, edges
    and faces, and 12, 30 and 20 vertices, edges and faces.
    
    We say that any point in an exceptional orbit is an {\em exceptional point.}  
    The Euler characteristic of $\Sigma$ (given by $2k(2-k)$) minus its exceptional
    points is divisible by $|G|$ since it admits a free $G$ 
    action:
    \begin{equation}  \label{eq:rest1}
	|G|\ |\ 2k(2-k)-\# \{ \rm exceptional\ points\}.
    \end{equation}
    Any point of an exceptional orbit of $G$ is fixed by some
    element of $G$ and hence by Lemma~\ref{th:fixedpts} it lies in the
    diagonal $\bp^1_\Delta\subset Z$.  Since we assume 
    $\Sigma$ to be smooth, it 
    cannot contain the component $\bp^1_\Delta$ (except in the non-reduced
    case $\Sigma=k\bp^1_\Delta$), so 
    \[
	\Sigma\cdot\bp^1_\Delta=(k,k)\cdot(1,1)=2k=\#\{{\rm exceptional\  points}\}+\ell|G| 
    \]
    where the right-hand side consists of the exceptional orbits and
    of $\ell \geq 0$ orbits of size $|G|$.  Thus
    \begin{equation}  \label{eq:rest2}
    |G|\ |\ 2k(1-k) 
    \end{equation}
    and we immediately deduce that $k=0$ or $1$ mod 3, 4 or 5 in the 
    respective tetrahedral, octahedral and icosahedral cases.  
    
    Since the quotient $\Sigma/G$ has genus one, 
    the Euler characteristic gives 
    \begin{equation}  \label{eq:rest3}
    \frac{2k(2-k)-\# \{ \rm exceptional\ points\}}{|G|}=-\#\{{\rm exceptional\  orbits}\}\geq -3.
    \end{equation}
       
    In the tetrahedral case, $|G|=12$, and if $k\geq 5$, then the
    left-hand side of (\ref{eq:rest3}) is less than $-3$ 
    which contradicts the inequality, so for $1<k<5$, $k=3$ and 4 give solutions of (\ref{eq:rest2}).  The exceptional orbits consist of 6 points in the $k=3$ case, and 4 plus 4 points in the $k=4$ case to give a solution of (\ref{eq:rest3}).
    In the octahedral case, $|G|=24$, so (\ref{eq:rest3}) implies that $k<8$.  Solutions of (\ref{eq:rest2}) must be 0 or 1 mod 4, hence $k=4$ or 5 and only 4 gives a solution of (\ref{eq:rest2}), with exceptional orbit of 8 points.  In the icosahedral case, $|G|=60$, so (\ref{eq:rest3}) implies that $k<11$, and together with the mod 5 condition we need only check $k=5$, 6 or 10.  Both $k=6$ and $k=10$ give solutions of (\ref{eq:rest3}), however only $k=6$ with exceptional orbit of 12 points allows a solution of (\ref{eq:rest3})  and the proposition is proven.
\end{proof}
    
It is useful to describe each of the exceptional $G$-orbits on $\bp^1_\Delta$ 
as the zeroes of a binary form --- a homogeneous polynomial in the two homogeneous coordinates for $\bp^1_\Delta$.  One then obtains three forms 
$K_v, K_e, K_f \in \bc[\zeta_0,\zeta_1]^{\rm hom}$ describing the positions
of the vertices, (midpoints of) edges and (midpoints of) faces of the
corresponding polyhedron. 
$G$ acts on $\bp^{1}_\Delta$ and this action can be transferred to the
vector spaces of binary forms of each degree.  
By construction, $K_{e}$, $K_{f}$ and $K_{v}$
are projectively invariant under $G$, and for each of these forms the
scalar factors under elements of $G$ give an abelian character of the
platonic group.  Since $G$ is finite, suitable products of these three
forms must be strictly invariant. 

In his famous book \cite{KleI} of
1884 on Galois theory, Felix Klein described the ring of $G$-invariant
forms for the three platonic groups; in particular, the unique monic
elements of minimal positive degree can be given in each case as
follows, in an obvious orientation of the polyhedra:
\begin{itemize}
    \item
    For $G= {A}_4$, $K_{e}(\zeta_0,\zeta_1)=\zeta_0 \zeta_1
    (\zeta_1^4-\zeta_0^4)$;
    \item
    For $G= {S}_4$, $
    K_{f}(\zeta_0,\zeta_1)=\zeta_1^8+14\zeta_0^4\zeta_1^4+\zeta_0^8$;
    \item
    For $G= {A}_5$, $
    K_{v}(\zeta_0,\zeta_1)=\zeta_0\zeta_1(\zeta_1^{10}
    +11\zeta_0^5\zeta_1^5-\zeta_0^{10})$; in this case, all
    projectively invariant forms are strictly invariant, a consequence
    of simplicity.
\end{itemize}
Given the result of Lemma~\ref{th:fixedpts}, candidates for smooth
$G$-invariant $(k,k)$ curves $\Sigma$ of lowest $k$ can now be written
as
\begin{equation}\label{Ansatzsym}
    (w-z)^k + \alpha \; \tilde{K} (w,z)=0,
\end{equation}
where $\tilde{K}(w,z)\in \bc[w,z]$ can be polarised to a $G$-invariant
element of $(\bc[w_0,w_1] \otimes \bc[z_0,z_1])^{\rm hom}$
that projects to the $G$-invariant forms $K_{e,f,v}$ above under the
dual $\iota^{*}$ of the inclusion $\iota: \bp^1_{\Delta}
\hookrightarrow \bpxp$.  The value of $k$ corresponds to
half of the degree of Klein's invariant form on $\bp^1_{\Delta}$
(i.e., $k=3$, $4$ or $6$), since $\iota^{*}$ projects $(p,q)$-forms
onto $(p+q)$-forms. Thus for each $G$, this procedure yields the symmetric curves of lowest $k$ in Proposition~\ref{mincurves}.

One way to find the appropriate function $\tilde{K}$ on $\bpxp$ from
Klein's invariant forms on $\bp^1_{\Delta}$ is to start with a
polynomial Ansatz incorporating invariance under $\sigma_{+}$ in
(\ref{sigma+}) and impose $G$-invariance, as illustrated in the
following example.  A reason why we should give $\sigma_{+}$ the
chance of being an extra symmetry of $\tilde{K}$ is that the parity
transformation ${\cal P}_{(0,0,1)}$ in (\ref{parity}) leaves the
zero-set of the particular Klein form associated with each platonic
group invariant; notice that this is not true in general for the other
two forms associated with a given platonic solid.

\begin{example}\label{invsfromKlein}
    Let $G= {A}_4$.  In terms of the inhomogeneous coordinate
    $\zeta:=\zeta_1/\zeta_0$, Klein's form $K_v(\zeta_0,\zeta_1)$
    corresponds to the polynomial $\zeta (\zeta^4 -1)$.  The
    projection $\iota^{*}$ is described by its action on generators
    $\iota^{*}(w)=\iota^{*}(z)=\zeta$, thus we write
    \[
    \textstyle
    \tilde{K}_v(w,z)=\left(\frac{w+z}{2}\right)\left(c_1(wz)^2+c_2(wz)\left(
    \frac{w+z}{2}\right)^2 +(1-c_1-c_2)\left(\frac{w+z}{2}
    \right)^4-1\right).
    \]
    In the orientation chosen, $ {A}_4$ is generated by the
    rotations
    \begin{equation}  \label{eq:gens}
	(w,z)\mapsto(-w,-z),\ \
	(w,z)\mapsto\left(\frac{1}{w},\frac{1}{z}\right) ,\ \ (w,z)\mapsto
	\left(\frac{w-i}{w+i},\frac{z-i}{z+i} \right).
    \end{equation}
    Imposing invariance under these, one finds $c_1=1$ and $c_2=0$. 
    Thus we write the Ansatz (\ref{Ansatzsym}) as
    \begin{equation} \label{tetrahedral}
	(w-z)^3+i\alpha \,(w+z)((wz)^2-1)=0.
    \end{equation}
    An easy check shows that this curve is real with respect to
    (\ref{reality}) if and only if $\alpha \in \br$.
\end{example}

An easy check shows that the curve (\ref{tetrahedral}) is real with respect to
(\ref{reality}) if and only if $a \in \br$.
The argument we used in section~\ref{sec:2nullar} shows that $m=0$
corresponds to $\alpha=\sqrt{3}$, given that the rational map of degree 3
with the $ {A}_4$-symmetry that we are using is~\cite{ManSut}
\[
R(z)=\frac{1-i\sqrt{3}z^2}{i\sqrt{3}z-z^3}.
\]
Moreover, $m\rightarrow \infty$ should correspond to $\alpha=0$ 
(as the limit spectral curve will be $\rm{SO}(3)$-symmetric, thus three
copies of the diagonal $\mathbb{P}^{1}_{\Delta}$).  Thus
in fact we can take
\[
0<\alpha < \sqrt{3}
\]
and it is easy to show that (\ref{tetrahedral}) is smooth and irreducible
for all these values of $\alpha$.

Using the same technique, we can find the corresponding polynomial
$\tilde{K}_f$ for $G= {S}_4$ and write (\ref{Ansatzsym}) as
\begin{equation} \label{octahedral}
    (w-z)^4+\alpha\,(w^4 z^4 + 6w^2z^2 +4wz(w^2+z^2)+1)=0.
\end{equation}
This curve lies in the 2-dimensional family of $ {A}_4$-symmetric
degree $(4,4)$ curves
\begin{equation} \label{tetrahedral4}
    (w-z)^4+\alpha\,(w^4 z^4 + 6w^2z^2 +4wz(w^2+z^2)+1)+
i\beta (w-z)(w+z)((wz)^2-1)=0
\end{equation}
which are real provided $\alpha,\beta \in \br$. It is easy to check that  
(\ref{tetrahedral4}) are all invariant under the rotations 
(\ref{eq:gens}), and if $\beta=0$ also under the extra order four rotation
\begin{equation}\label{extraS4}
(w,z)\mapsto (iw,iz).
\end{equation}
In fact, the curves (\ref{tetrahedral4}) are still invariant under 
\begin{equation}\label{hiddenS4}
(w,z) \mapsto (iz,iw),
\end{equation}
which together with (\ref{eq:gens}) generates $ {S}_4$, and then
$\Sigma/ {S}_4$ is also an elliptic curve, double-covered by
$\Sigma/ {A}_4$; however,
since (\ref{hiddenS4}) is not a rotation we do not call this 
$ {S}_4$ an octahedral
symmetry group. Note that the curves (\ref{tetrahedral}) in the 
$k=3$ case are also invariant under this hidden $ {S}_4$ symmetry,
but in this case $\Sigma/ {S}_4$ is a rational curve. 
As for $k=3$, the limit $m\rightarrow \infty$ of (\ref{tetrahedral}) 
should give $\alpha=0$, and the
value of $\alpha$ for $m=0$ can be calculated from the $S_4$-invariant 
rational map~\cite{ManSut}
\[
R(z)=\frac{z^4+2\sqrt{3}i z^2+1}{z^4-2\sqrt{3}i z^2+1}
\]
to be $\alpha=1$, thus we take
\[
0<\alpha< 1;
\]
again, $\Sigma$ is smooth and irreducible for all these values of $\alpha$.

For $G= {A}_5$ we have one parameter $\alpha \in \br$ in
\begin{eqnarray} \label{icosahedral}
    &(w-z)^6+\alpha\,(9(w^6+z^6)+9wz(w^4+z^4)+10(wz)^2(w^2+z^2)+10(wz)^3&
    \nonumber \\
    &+3(w+z)((wz)^5-1))=0.&
\end{eqnarray}
This final example fails the vanishing cohomology condition~3.\ (cf.\ Propositions~\ref{th:vancoh} and \ref{th:vancoh2}).  As in the euclidean case, it is necessary to multiply the degree 6 polynomial by $(w-z)$ to obtain a reducible degree $(7,7)$ curve, so that the holomorphic sections on the degree $(6,6)$ component have to satisfy enough further conditions to be forced to vanish.  We will not treat this reducible spectral curve here.

\section{Tetrahedral and octahedral symmetry}\label{sec:tetrahedral} 

In the rest of the paper, we shall relate the symmetric $(k,k)$ curves above 
to spectral curves of hyperbolic monopoles with a given mass. 
Our main aim is to obtain
the mass associated with these curves, and the basic strategy will consist of transferring calculations on $\Sigma$ to the quotients by
their platonic symmetry groups,
\begin{equation}\label{elliptquot}
      \pi: \Sigma \rightarrow \Sigma/G =: E.
\end{equation}
In this way, the complex analysis needed to relate the 
monopole mass $m$ to the parameter $\alpha$ in the Ans\"atze will have 
the same flavour as the $k=2$ discussion above. 
However, unlike the $2$-monopole case, we will now have to deal with a nontrivial condition~3., which
we shall approach using some classical algebraic geometry.
It will turn out that condition 3.\ will now 
also play a r\^ole in the mass calculation itself.
Our results can be summarised as follows:

\begin{theorem} \label{th:3A4}
    There is a unique ${\rm PSL}_{2}\bc$-orbit of tetrahedrally
    symmetric 3-monopoles with mass $m>0$; a representative is the
    centred monopole with spectral curve (\ref{tetrahedral}), where
    $\alpha$ and $m$ satisfy the relation
    \begin{equation}\label{relamA4}
	\wp\left(\left.\frac{2 \varpi_1}{2m+3}\right|\varpi_1,\varpi_2\right)
    =\frac{1}{12}-\frac{1}{\alpha^2}
    \end{equation}
    involving the Weierstra\ss\ $\wp$-function of the elliptic 
    curve with invariants (\ref{g_23A4}).    
\end{theorem}

\begin{theorem} \label{th:4A4}
    There is a unique ${\rm PSL}_{2}\bc$-orbit of
    octahedrally symmetric 4-monopoles with representative
    (\ref{octahedral}), whose mass is determined by the relation
    \begin{equation}\label{relamS4}
	\wp\left(\left.\frac{3 \varpi_1}{m+2}\right|\varpi_1,\varpi_2\right)
    = \frac{-4\alpha^4+10\alpha^3-115\alpha^2+60\alpha-3}
      {54 \alpha^2 (\alpha + 1)^2},
    \end{equation}
    where the Weierstra\ss\  $\wp$-function has invariants (\ref{g_2S4}) and
    (\ref{g_3S4}).
\end{theorem}

In both cases, we actually construct a curve together with a linear flow 
on its jacobian variety that avoids the theta-divisor; 
this will be discussed in
sections \ref{sec:canemb} and \ref{sec:thdiv}.  By quite general results (see \cite{AdlMoe,GriLin,HitMG}) this gives rise to a Lax system which we refer to as a monopole.

We express the relation between $\alpha$ and $m$ for each Ansatz in terms
of the 1-parameter family of elliptic curves $E\equiv E_\alpha$; thus the
half-periods $\varpi_1$ and $\varpi_2$ in equations (\ref{relamA4}) and 
(\ref{relamS4}), which we shall specify carefully below, are functions of $\alpha$. To obtain this relation, our starting point is a 
reciprocity argument similar to the one in section~\ref{sec:reciproc}.
We let $\xi_j$ denote trivialising sections of $L^{2m+k}|_{\Sigma}$ over
the open sets $U_j\cap \Sigma$ ($j=1,2$) defined by~(\ref{U1U2}), and use them to
obtain differentials of the third kind 
$\omega_j={\rm d} \log \xi_j$ on $\Sigma$, related through
\[
\omega_{1}=(2m+k)\left(\frac{{\rm d}w}{w}-\frac{{\rm d}z}{z} \right) +\omega_2.
\]
We want to apply the reciprocity law generalising (\ref{reciprocity})
to $\omega_1$ and to $\pi^{*}\omega$, where $\omega$ is a global
holomorphic 1-form on $E$;
obviously, $\pi^{*}\omega$ is a differential of the first kind on
$\Sigma$ which is invariant under the $G$-action.  Fix a
canonical basis $\{ a_\ell, b_\ell \}_{\ell=1}^{g}$ for
$H_1(\Sigma;\bz)\cong \bz^{2g}$, where $g=(k-1)^2$, 
and define integers $m_\ell, n_\ell$ by
\[
m_\ell:=\frac{1}{2\pi i}\oint_{b_\ell}\omega_1, \qquad
n_\ell:=-\frac{1}{2\pi i}\oint_{a_\ell}\omega_1.
\]
We denote by $p_j, q_j$ the poles of
$\frac{{\rm d}w}{w}-\frac{{\rm d}z}{z}$; they can be found explicitly using
the equation for $\Sigma$. Our conventions are such that the residues of $\omega_1$ at these poles are given by
\[
{\rm Res}_{p_j}(\omega_1)=-(2m+k) \quad \mbox{and} \quad
{\rm Res}_{q_j}(\omega_1)=2m+k \quad j=1,\ldots, 2(k-1).
\]
Using the explicit description in section~\ref{sec:genmconstr}, we shall
see that in both cases the points $p_j$ and $q_j$ lie on two separate $G$-orbits, and we set
\[
p:=\pi(p_j),\qquad q:=\pi(q_j).
\] 
The reciprocity law can now be written as
\begin{equation} \label{reciprk3A4} 
    \sum_{\ell=1}^{g} \left|\begin{array}{cc}
    \oint_{a_\ell} \pi^{*}\omega &  -n_\ell \\
    \oint_{b_\ell} \pi^{*}\omega &  m_\ell 
    \end{array}\right| = (2m+k) \sum_{j=1}^{2(k-1)} 
    \int_{p_j}^{q_j}\pi^{*}\omega,
\end{equation}
where the paths of integration in the right-hand side avoid the
1-homology basis.  

Let $I=]0,\sqrt{3}[$ if $k=3$, respectively $I=]0,1[$ if $k=4$, denote 
the range of $\alpha$.
Equation (\ref{reciprk3A4}) determines $\alpha\in I$ as a function of $m$, once
the integers $m_\ell, n_\ell$ are known. In fact, these integers are
constant along the isotopy of curves $\Sigma=\Sigma_\alpha$ given by (\ref{tetrahedral}) or (\ref{octahedral}):

\begin{lemma} \label{nojump}
The integers $m_\ell, n_\ell$ defined by equation (\ref{reciprk3A4})
are independent of $\alpha \in I$.
\end{lemma}
\begin{proof}
Suppose that $\alpha$ and $m$ are fixed.
For each element $\omega_i$ of a basis of global holomorphic
1-forms on $\Sigma$, a reciprocity relation like (\ref{reciprk3A4}) can be written,
where $\pi^{*}\omega$ is replaced by $\omega_i$. Taking real and imaginary 
parts of these $g$ equations, one obtains $2g$ real equations in the $2g$ 
real unknowns $m_\ell, n_\ell$. If the basis $\{\omega_i \}_i$ is dual
to the 1-cycles $a_\ell$, then we see that the equations corresponding to 
the imaginary parts decouple the variables $n_\ell$, and we can solve for
them as the matrix of coefficients is nonsingular by the second Riemann bilinear relations~\cite{GriHar}. Substitution in the equations corresponding to the
real parts give immediately the $m_\ell$, as the matrix of coefficients is
then the identity by construction. We conclude that we can always solve for
$m_\ell, n_\ell$, and the solutions are given as linear fractional functions
of the periods of $\Sigma$ and the real and imaginary parts of the integrals 
$\int_{q_j}^{p_j}\omega_i$.

We now argue that the periods of $\Sigma$ are continuous functions of
$\alpha$. Let $\{ \varphi_\alpha\}_{\alpha\in I}$ be an isotopy of $Z$ such
that $\varphi_\alpha(\Sigma_{1/2})=\Sigma_\alpha$ for all $\alpha \in I$, say. Fix representatives of a basis of 1-cycles $\{c_\ell\}_\ell$ on $\Sigma_{1/2}$, and transfer them to each $\Sigma_\alpha$ using 
$\varphi_\alpha$. Moreover, equip each curve with a basis of global holomorphic
1-forms $\omega_{i}^{\alpha}$ obtained by adjunction (i.e.\ taking Poincar\'e\ residues of a fixed set of $g$ holomorphic 2-forms on a neighbourhood of the family of curves in $Z$); this is possible as each $\Sigma_\alpha$ is smooth. Then one finds
\[
\oint_{\varphi_{\alpha*}c_\ell}\omega_{i}^{\alpha}=
\oint_{c_\ell}\varphi_\alpha^{*}\omega_{i}^{\alpha}
\]
and the right-hand side is an integral of a continuous function 
of $\alpha$, which is itself continuous in $\alpha$. 
Clearly, the same type of argument
works for the paths connecting poles of $\frac{dw}{w}$ to poles of 
$\frac{dz}{z}$, and the lemma follows.
\end{proof}

To transfer the calculation to $E$, we define
\begin{equation} \label{cA4}
    \textstyle c:=\pi_{*} \left(\sum_{j=1}^{g}(m_j a_j + n_j b_j)
    \right) =:\ell_1 a + \ell_2 b,
\end{equation}
where $\{a,b\}$ is a standard basis of $H_1(E;\bz)$, and obtain
\begin{equation} \label{reciprEA4}
    \oint_c \omega = 2 (2m+k)\int_{p}^{q} \omega
\end{equation}
after using $\pi$ to change variables and clearing a factor of $\deg
\pi=|G|$.
As a consequence of Lemma~\ref{nojump}, the components of $c$ will 
not change in an isotopy of bases of $H_1(E;\bz)$ defined for $\alpha \in A$.
Note that, although the paths connecting $q_j$ to $p_j$ on $\Sigma$ had
to be chosen to have zero intersection with the elements of
$H_1(\Sigma;\bz) \hookrightarrow H_1(\Sigma,\{p_j,q_j\};\bz)$, this
does not necessarily hold as we project to $E$. However, there is a simple
criterion to test
whether a path $\gamma$ from $q$ to $p$ on $E$ is the image of a suitable path
on $\Sigma$, namely, the components of the 1-cycle $c \in H_1(E; \bz)$ 
must be independent of $m$ for solutions 
$(\alpha,m)$ of (\ref{reciprEA4}); once
at least two such solutions are obtained, $\gamma$ can be found 
systematically by expanding in a basis of
$H_1(E,\{ p,q \};\bz)\cong \bz^3$.
In section~\ref{sec:halfintm}, we shall describe a 
general procedure that allows one to obtain data $(\alpha,m)$ for 
half-integer values of $m$. From these, we
will be able to calculate both a suitable
path $\gamma$ and the 1-cycle $c$ from an explicit construction of $E$
in section~\ref{sec:genmconstr}. Finally, we will use uniformisation on
$E$ to evaluate the mass constraint (\ref{reciprEA4}) in terms of elliptic 
functions.

\section{Canonical embedding} \label{sec:canemb}

The canonical embedding of a genus $g$ curve $\Sigma$ is a map
\[
\Sigma\to\bp^{g-1}
\]
defined by $z\mapsto[\omega_1(z):\ldots:\omega_g(z)]$ with respect to a
local trivialisation of the canonical line bundle $K_{\Sigma}$.  This
map is an embedding except when the curve is hyperelliptic, in which
case it maps 2-to-1 onto a rational curve in $\bp^{g-1}$.  It gives a
useful geometric version of Riemann--Roch which we will use.

By adjunction, the canonical embedding of a curve embedded in a surface $X$ can be induced by an embedding $X\to\bp^{g-1}$.  We will use this to give an explicit description of the canonical embedding of a curve in the
quadric $Q:=\bpxp$.  (This can be done similarly for $\bp^2$.  An immediate consequence is that no smooth hyperelliptic curve with $g>1$ embeds in $\bp^2$ or $\bpxp$.)

\begin{lemma}    \label{th:canem}
    For $k>2$, the canonical embedding of a smooth $(k,k)$ curve
    $\Sigma\subset\bpxp$ is the composition
    \begin{eqnarray*}
	\Sigma\hookrightarrow\bpxp& \!\!\hookrightarrow\!\!&\bp^{k(k-2)}\\
	(w,z)&\!\!\mapsto\!\!&[1:w:...:w^{k-2}:z:wz:...:w^{k-2}z:...:
	z^{k-2}:wz^{k-2}:...:w^{k-2}z^{k-2}]
    \end{eqnarray*}
\end{lemma}
\begin{proof}
    As mentioned above we will use the adjunction formula
    $K_{\Sigma}=\mathcal{O}(k-2,k-2)|_{\Sigma}$.  Sections of $K_{\Sigma}$ can be
    identified with sections that extend to all of $Q=\bpxp$ since the
    relation introduced by the equation of $\Sigma$ is in higher
    degree than $k-2$, or equivalently the outer two cohomology groups
    vanish in the following exact sequence:
    \[ 
    H^0(Q,\mathcal{O}(-2,-2))\to H^0(Q,\mathcal{O}(k-2,k-2))\to
    H^0(\Sigma,\mathcal{O}(k-2,k-2))\to H^1(Q,\mathcal{O}(-2,-2)).
    \]
    The polynomials $1,w,\ldots,w^{k-2},z,wz,\ldots,w^{k-2}z,\ldots,
    z^{k-2},wz^{k-2},\ldots,w^{k-2}z^{k-2}$ give a basis of the space 
    of sections of $\mathcal{O}(k-2,k-2)$, where we are using $U_1$ in
    (\ref{U1U2}) as a local trivialising set.
\end{proof}

\begin{lemma}   \label{th:lineqrel}
    For any smooth $(3,3)$ curve $\Sigma\subset \bpxp$, a
    nontrivial linear equivalence relation 
    \[ 
    p_1+p_2+p_3\sim q_1+q_2+q_3
    \] 
    is equivalent to $p_1+p_2+p_3\sim\mathcal{O}_\Sigma(1,0)$ or
    $\mathcal{O}_\Sigma(0,1)$.
\end{lemma}
\begin{proof}
    This is a well-known fact.  Clearly any two fibres of the projection of $\bpxp$ to the first
    factor are linearly equivalent, and the same for projection to the
    second factor.  It ends up that these are the only linear
    equivalences between triples of points on a smooth degree $(3,3)$
    curve in $\bpxp$.   An equivalent statement is that a $g^1_3$ on $\Sigma$ must be 
    $\mathcal{O}_\Sigma(1,0)$ or $\mathcal{O}_\Sigma(0,1)$.  
    (Recall that a $g^r_d$ is any linear system of degree $d$ and dimension
    $r+1$, see for example~\cite{ACGHGe}.)  We include the proof as a 
    warm-up for similar results.

    We use the canonical embedding 
    \[
    \phi:\Sigma\to\bp^3
    \]
    to prove this.  The images $\phi(p_1)$, $\phi(p_2)$ and
    $\phi(p_3)$ lie inside a $\bp^2\subset\bp^3$.  When there exists
    the nontrivial relation $p_1+p_2+p_3\sim q_1+q_2+q_3$, the
    geometric version of Riemann--Roch says that $\phi(p_1)$,
    $\phi(p_2)$ and $\phi(p_3)$ lie inside a $\bp^1\subset\bp^3$.  
    
    Suppose two of the points lie on a $(1,0)$ or $(0,1)$ curve,
    say $p_1$ and $p_2$ lie on a $(1,0)$ curve.  Then their images
    determine a line in $\bp^3$, $[1:w:c:wc]$ for some constant $c$.
    The image of $p_3$, given by $[1:w:z:wz]$, lies on the curve only
    if $z=c$, i.e. $p_3$ lies on the same $(1,0)$ curve, proving the
    lemma.
    
    If no two points lie on a $(1,0)$ or $(0,1)$ curve, then the three
    points lie on a smooth $(1,1)$ curve.  Any smooth $(1,1)$ curve is
    equivalent to the diagonal $\bp^1_\Delta\subset\bpxp$ which has images
    $[1:w:w:w^2]$ inside $\bp^3$, and any three different points of
    this form are linearly independent due to the $1,w,w^2$ terms.  This
    contradicts the fact that the points $p_1$, $p_2$ and $p_3$ span
    $\bp^1$.  Essentially the same argument is used when some of the
    $p_i$ coincide, so the result follows.
\end{proof}
    
    Recall that a divisor $D$ is {\em special} if $H^1(D)\neq 0$, or equivalently if it has more sections than a generic divisor of the same degree.
    \begin{lemma}  \label{th:superab}
    A positive divisor $D$ of degree at most 7 on a $(4,4)$ curve $\Sigma\subset Q$ is special precisely when one of the following holds:
    
    (a) $D$ contains four points on a $(1,0)$ or $(0,1)$ curve;
    
    (b) $D$ contains six points on a $(1,1)$ curve.
    
    \end{lemma}
    \begin{proof}
    The proof of this requires a systematic analysis of many separate
    cases.  To avoid this we will instead refer to an exercise from \cite{ACGHGe}, p.~199, stating that any collection of at most seven points in $\bp^3$ that fails to impose independent conditions on quadrics contains one of the following: (i) four collinear points; (ii) six points on a conic; (iii) seven coplanar points.
    
    Lemma~\ref{th:canem} shows that the canonical embedding of $\Sigma$ factors through the quadric $Q$.  More is true.  It also factors through $\bp^3$, so we have 
    \[
    \Sigma\hookrightarrow Q\hookrightarrow\bp^3\hookrightarrow\bp^8
    \]
    with rightmost map $[z_0:z_1:z_2:z_3]\mapsto
[z_0^2:z_0z_1:z_0z_2:z_0z_3:z_1^2:z_1z_3:z_2^2:z_2z_3:z_3^2]$ the degree two Veronese map, where the monomial $z_1z_2$ is missing since it is equal to $z_0z_3$ on $Q$.  Thus a collection of points on $\Sigma$ representing a positive divisor $D$ gives a collection of points in $\bp^3$.  By the geometric version of Riemann--Roch, $D$ is special when the images of the points in $\bp^8$ are dependent, so one of (i), (ii), or (iii) occurs.   
    
    The intersection of a line $L\subset\bp^3$ and $Q$ either consists of two points (counted with multiplicity) or $L\subset Q$.  If (i) occurs then four points from $Q$ lie on a line $L\subset\bp^3$ and hence $L\subset Q$.  The only such lines are $(1,0)$ or $(0,1)$ curves so case (a) holds.
    
    The intersection of a conic $C\subset\bp^3$ and $Q$ either consists of four points or $C\subset Q$.   If (ii) occurs, then six points from $Q$ lie on a conic $C\subset\bp^3$ and hence $C\subset Q$.  Conics in $Q$ are $(1,1)$ curves so (b) holds.  The intersection of a plane and the quadric in $\bp^3$ is a conic in $\bp^2$ and a $(1,1)$ curve in $Q$, so if (iii) occurs then seven points lie on a $(1,1)$ curve and again (b) holds.
 \end{proof}

\begin{lemma}   \label{th:lineqrel1}
    For any smooth $(4,4)$ curve $\Sigma\subset \bpxp$, a
    nontrivial linear equivalence relation
    \[ 
    p_1+p_2+p_3+p_4\sim q_1+q_2+q_3+q_4
    \]
    implies $p_1+p_2+p_3+p_4\sim\mathcal{O}_\Sigma(1,0)$ or
    $\mathcal{O}_\Sigma(0,1).$
\end{lemma}
\begin{proof}
    This is immediate from Lemma~\ref{th:superab}.
\end{proof}

\begin{lemma}   \label{th:lineqrel2}
    For any smooth $(4,4)$ curve $\Sigma\subset \bpxp$, the existence of two
    independent linear equivalence relations
    \[ 
    p_1+...+p_8\sim q_1+...+q_8\sim r_1+...+r_8
    \]
    implies that there exists $p_9,...,p_{12}$ such that $p_1+...+p_8+p_9+...+p_{12}\sim\mathcal{O}_\Sigma(2,1)$ or
   $ \mathcal{O}_\Sigma(1,2)$.
\end{lemma}
\begin{proof}
By {\em independent} relations we mean that the points satisfy one of the three equivalent conditions: $h^0(p_1+\ldots+p_8)>2$; or $p_1+\ldots+p_8$ defines a $g^2_8$; or the images of $p_1$, \ldots, $p_8$ in $\bp^8$ live in a $\bp^5$.  Since the images of $p_1$, \ldots, $p_8$ in $\bp^8$ satisfy two relations, any subset of seven points satisfies a relation and we can apply Lemma~\ref{th:superab} to all eight such subsets.

   (I) If seven points contain six points on a smooth $(1,1)$ curve, then 
no four points of the eight points are contained on a line, so all subsets of
seven points must contain six points in a conic.  Since a conic is determined by three points, the eight conics must coincide.  Thus the eight points consist of seven points on a smooth conic and an eighth general point.  In particular, the eight points lie inside a $(2,1)$ or $(1,2)$ curve as claimed.
   
   (II) If four of the eight points are contained on a line, say a $(1,0)$ curve, then since five points cannot lie on a line --- $\Sigma$ is a 
smooth $(4,4)$ curve ---, when a subset of seven points does not include these four points, it must include either four points on another line, or three points on a $(0,1)$ curve so that it has $3+3$ points on a reducible $(1,1)$ curve.  In both these cases, seven points lie on two lines, and the eighth point is general, so in particular the eight points lie inside a $(2,1)$ or $(1,2)$ curve as claimed.
   
   (III) If neither (I) nor (II), i.e.\ no four points lie on a  line and 
no six points lie on a smooth $(1,1)$ curve then case (b) of Lemma~\ref{th:superab} must apply to all eight subsets, with a reducible $(1,1)$ curve, i.e.\ $3+3$ points must lie on  a $(1,0)$ and $(0,1)$ curve.  A subset of 
seven points may not contain the three points on the $(1,0)$ curve, so another $(1,0)$ curve must contain three points, and this takes at least two extra points from the eight points, i.e.\ $3+2+3$ points distributed on a $(1,0)$, $(1,0)$ and $(0,1)$ curve.  Similarly we also need another $(0,1)$ curve 
with three points and this requires at least one more point, so nine points are required.  Thus, case (III) is empty and the lemma is proven.
\end{proof}

\section{Crossing the theta-divisor} \label{sec:thdiv}

\subsection{$k=3$ tetrahedral symmetry}

Consider the smooth genus four curve $\Sigma\subset\bpxp$ with equation
(\ref{tetrahedral}) for $\alpha\in\mathbb{R}^+$.  The bundle
$L^s(1,0)|_{\Sigma}$ is a degree 3 bundle, or equivalently a degree 3
divisor on $\Sigma$.  Recall that the divisor $D$ of
$L^s(1,0)|_{\Sigma}$ meets the theta-divisor if one of the following
equivalent properties holds:
\begin{enumerate}
    \item 
    $D\sim p_1+p_2+p_3$\ \ i.e.\ $D$ is linearly equivalent to a
    positive divisor; 
    \item 
    there exists a meromorphic function $f$ on $\Sigma$ such that
    $(f)+D\geq 0$;
    \item 
    there exists a nontrivial holomorphic section of the line bundle
    $L^s(1,0)|_\Sigma$;
    \item
    $H^0(\Sigma,L^s(1,0))\neq 0$.
\end{enumerate}
Here we prove that, when $0<s<2m+2$, the divisor of
$L^s(1,0)|_{\Sigma}$ {\em does not} meet the theta-divisor.

\begin{proposition}  \label{th:vancoh}
    $H^0(\Sigma,L^s(1,0))=0$ for $0<s<2m+2$.
\end{proposition}
\begin{proof}
    The geometry implies that $L$ is invariant under the $A_4$-action,
    since $L$ comes from the standard ${\rm U}(1)$ monopole which is
    symmetric under ${\rm SO}(3)$.  In other words, $\mathcal{O}(1,-1)$ is
    invariant under the ${\rm SO}(3)$ action on $\bpxp$ so
    $L_\Sigma:=\mathcal{O}(1,-1)|_{\Sigma}$ is invariant under $A_4$.  In
    particular, $L_\Sigma$ is the pull-back of a degree zero line bundle
    $\hat{L}$ on $E=\Sigma/A_4$, and $L_\Sigma^s=\pi^*\hat{L}^s$ for
    $s\in\mathbb{C}$.  Since $E$ is an elliptic curve, as a divisor
    $\hat{L}^s\sim p-p'$, and all such divisors are represented as $s$
    varies over the complex numbers.  Thus, since $\hat{L}^s$ pulls
    back to $L_\Sigma^s$, we can represent the divisor of any $L_\Sigma^s$ by
    \[ 
    L_\Sigma^s\sim {\rm Orb}_p-{\rm Orb}_{p'}
    \] 
    where ${\rm Orb}_p$ is the $A_4$-orbit in $\Sigma$ that lies over
    $p\in E$.  So questions involving $L_\Sigma^s$ are questions about orbits
    of $A_4$.
    
    We will show that the difference of two orbits of $A_4$ in
    $\Sigma$ plus the divisor class $\mathcal{O}_\Sigma (1,0)$ lies in the 
    theta-divisor,
    \begin{equation}  \label{eq:lineq}
	{\rm Orb}_p-{\rm Orb}_{p'}+\mathcal{O}_\Sigma(1,0)\sim q_1+q_2+q_3,
    \end{equation}
    in the following trivial cases: 
    \[ 
    {\rm Orb}_p-{\rm Orb}_{p'}\sim 0\ \ \ {\rm or\ }\ \ {\rm Orb}_p-{\rm Orb}_{p'}\sim
    L_\Sigma^{-1}=\mathcal{O}_\Sigma (-1,1).
    \]
    The action of $A_4$ on the left hand side of (\ref{eq:lineq})
    preserves the two orbits and also preserves the linear equivalence
    class $\mathcal{O}_\Sigma(1,0)$ since the action, given in
    (\ref{eq:gens}), preserves the two $\bp^1$ factors of $\bpxp$. 
    Thus, for any $g\in A_4$,
    \begin{equation}  \label{eq:lineqrel}
	q_1+q_2+q_3\sim g q_1+g q_2+g q_3.
    \end{equation}
    The collection $\{q_1,q_2,q_3\}$ cannot be invariant under $A_4$,
    since the orbits of $A_4$ have size 12 and 6.  Thus we can choose
    a $g\in A_4$ that does not preserve $\{q_1,q_2,q_3\}$, so the
    linear equivalence relation (\ref{eq:lineqrel}) is a nontrivial
    relation between degree 3 positive divisors.  By
    Lemma~\ref{th:lineqrel}, this implies one of the two cases
    \[ 
    q_1+q_2+q_3\sim\mathcal{O}_\Sigma(1,0)\ \ \ {\rm or\ }\ \
    q_1+q_2+q_3\sim\mathcal{O}_\Sigma(0,1).
    \]
    If $q_1+q_2+q_3\in\mathcal{O}_\Sigma(1,0)$, then (\ref{eq:lineq}) reduces
    to 
    \[ 
    {\rm Orb}_p-{\rm Orb}_{p'}\sim 0,
    \] 
    in other words $L_\Sigma^s$ is trivial, so $s=0$ (or a multiple of $2m+3$
    since $L_\Sigma^{2m+3}\cong\mathcal{O}_\Sigma$.)  If
    $q_1+q_2+q_3\in\mathcal{O}_\Sigma(0,1)$, then
    \[ 
    0\sim {\rm Orb}_p-{\rm Orb}_{p'}+\mathcal{O}_\Sigma(1,0)-(q_1+q_2+q_3)\sim L_\Sigma^{s+1},
    \] 
    so $s=-1$ (plus a multiple of $2m+3$.)  In particular, when
    $0<s<2m+2$, $L^s(1,0)|_\Sigma$ does not meet the theta-divisor.
\end{proof}

\subsection{$k=4$ octahedral symmetry}

Orbits of $S_4$ on $\Sigma$ consist of 24 points, except for
the one exceptional orbit of 8 points given by the $(1,1)$ divisor
$\bp^1_\Delta\cap\Sigma$.

\begin{proposition}  \label{th:vancoh2}
    $H^0(\Sigma,L^s(2,0))=0$ for $0<s<2m+2$.
\end{proposition}
\begin{proof}
    As before, we reduce questions involving $L_\Sigma^s$ to questions about
    orbits of $S_4$ using
    \[ 
    L_\Sigma^s\sim {\rm Orb}_p-{\rm Orb}_{p'}
    \]
    for $p,p'\in E=\Sigma/S_4$.

    The difference of two orbits of $S_4$ in $\Sigma$ plus a
    $\mathcal{O}_\Sigma(2,0)$ divisor lies in the theta-divisor,
    \begin{equation}  \label{eq:lineq2}
	{\rm Orb}_p-{\rm Orb}_{p'}+\mathcal{O}_\Sigma(2,0)\sim  q_1+q_2+...+q_8,
    \end{equation}
    in the following trivial cases:
    \begin{equation}  \label{eq:trivcases}
	{\rm Orb}_p-{\rm Orb}_{p'}\sim L_\Sigma^{\epsilon},\  \epsilon=0,-1\ {\rm or\  }-2.
    \end{equation}
 
    The action of $S_4$ on the left-hand side of
    (\ref{eq:lineq2}) preserves the two orbits and also preserves the
    linear equivalence class $\mathcal{O}(2,0)$.  Thus, for any $g\in S_4$,
    \begin{equation}  \label{eq:lineqrel2}
	q_1+q_2+...+q_8\sim g q_1+g q_2+...+gq_8.
    \end{equation}
	
    If the collection $\{q_1,q_2,...,q_8\}$ is invariant under $S_4$
    then it consists of the exceptional orbit and
    \[ 
    q_1+q_2+...+q_8\sim \mathcal{O}_\Sigma(1,1),
    \]
    which yields $\epsilon=-1$ in (\ref{eq:trivcases}).
	
    If the collection $\{q_1,q_2,...,q_8\}$ is not invariant under
    $S_4$, then we can choose a $g\in S_4$ that does not
    preserve $\{q_1,q_2,...,q_8\}$, so the linear equivalence relation
    (\ref{eq:lineqrel2}) is a nontrivial relation between degree 8
    positive divisors, or equivalently $\dim H^0(\Sigma,L^s(2,0))\geq
    2$.  We may assume that there is no section in
    $H^0(\Sigma,L^s(2,0))$ that is invariant under $S_4$, or
    more generally generates a 1-dimensional representation of
    $S_4$, since the zero set of such a section would have to
    be the exceptional orbit of $S_4$ and we get the previous
    case of $\epsilon=-1$.  Thus, if $H^0(\Sigma,L^s(2,0))=\bc^2$ it
    must be an irreducible representation of $S_4$.  The
    standard 2-dimensional representation is given by the action on
    $H^0(\Sigma,\mathcal{O}(1,0))$ and thus the tensor product
    $H^0(\Sigma,L^s(3,0))$ is the 4-dimensional representation with a
    1-dimensional irreducible component generated by a section
    $\xi=w\chi\in H^0(\Sigma,L^s(3,0))$ for $\chi\in
    H^0(\Sigma,L^s(2,0))$.  The zero set of $\xi$ is invariant under
    $S_4$ and contains a $(1,0)$ divisor.  But this is impossible
    simply because any orbit of $S_4$, or collection of exceptional
    orbits, contains at most three points in a $(1,0)$ curve.
    
    Thus $\dim H^0(\Sigma,L^s(2,0))> 2$, so by Lemma~\ref{th:lineqrel2}
    there exist points $q_9,\   q_{10},\  q_{11},\  q_{12}$ such that 
    \begin{equation}   \label{eq:12pt}
    q_1+q_2+...+q_{12}\sim\mathcal{O}_\Sigma(2,1)\  
    {\rm or\ } \mathcal{O}_\Sigma(1,2)
    \end{equation} 
    and (\ref{eq:lineq2}) becomes 
    \[
    {\rm Orb}_p-{\rm Orb}_{p'}+\mathcal{O}_\Sigma(2,0)+q_9+q_{10}+q_{11}+q_{12}\sim
    \mathcal{O}_\Sigma(2,1)\  {\rm or\ } \mathcal{O}_\Sigma(1,2).
    \] 
    Since ${\rm Orb}_p-{\rm Orb}_{p'}=L_\Sigma^s$ and any multiple of $\mathcal{O}_\Sigma(1,-1)$ is a power of $L_\Sigma$, we can adjust this expression and choose two new orbits $\tilde{p}$ and $\tilde{p}'$ so that
    \begin{equation}  \label{eq:lineq3}
	{\rm Orb}_{\tilde{p}}-{\rm Orb}_{\tilde{p}'}+\mathcal{O}_\Sigma(1,0)\sim  q_9+q_{10}+q_{11}+q_{12},
    \end{equation}
    which is a reduction of the original problem.  Again we use $g\in S_4$ to get a nontrivial linear equivalence $q_9+q_{10}+q_{11}+q_{12}\sim gq_9+gq_{10}+gq_{11}+gq_{12}$ and apply Lemma~\ref{th:lineqrel1} to obtain
    \[
    q_9+q_{10}+q_{11}+q_{12}=\mathcal{O}_\Sigma(1,0)\ {\rm
    or\ }\mathcal{O}_\Sigma(0,1).
    \] 
    Put this back into (\ref{eq:12pt}) to get
    \[
    q_1+q_2+...+q_8=\mathcal{O}_\Sigma(2,0),\  \mathcal{O}_\Sigma(1,1),\  {\rm
    or\ }\mathcal{O}_\Sigma(0,2)
    \] 
    and hence
    \[ 
    {\rm Orb}_{p}-{\rm Orb}_{p'}\sim L_\Sigma^{\epsilon}, \epsilon=0,-1\ {\rm or\  }-2.
    \] 
    In particular, we have proved that when $0<s<2m+2$, $L^s(2,0)|_\Sigma$ does not meet the theta-divisor.
\end{proof}

\noindent    
{\em Remark.} Hitchin \cite{HitPS} proves a result analogous to Proposition~\ref{th:vancoh2} in the euclidean case using a slightly different method.  His treatment of the tetrahedral case is essentially the same
as ours, relying on knowledge of $g^1_3$ divisors on a genus four curve.  Whereas we extend this approach  through an analysis of $g^1_4$ and $g^2_8$ divisors on a genus nine curve, Hitchin analyses the representation theory of $S_4$ more thoroughly and produces a beautiful application of the McKay correspondence.  We need a small amount of representation theory to 
exclude a two-dimensional representation of $S_4$, mainly because the space of $g^1_8$ divisors is too large to manage.  The family of tetrahedrally symmetric $(4,4)$ curves (\ref{tetrahedral4}) cannot be treated 
using the representation theory approach, while the method here 
does generalise.

\section{Half-integer mass}\label{sec:halfintm}
    
     We have shown that triviality of $L^{2m+k}$ on $\Sigma$ determines a
relation between $\alpha$ and $m$.  The vanishing of
$H^0(\Sigma,L^{2m}(k-2,0))$ also has a direct consequence on the
parameter $\alpha$ of the symmetric spectral curve.  Since 
\[
H^0(\Sigma,L^{2m}(k-2,0))= H^0(\Sigma,\mathcal{O}(-2,k)),
\] 
a section can be described by 
\[ 
\xi_1(w,z)=
w^{-2}z^k \xi_2 (w^{-1},z^{-1})+\psi(w,z) \chi(w^{-1},w,z^{-1},z),
\] 
where $\xi_{1}, \xi_2$ give the section over the two coordinate patches
$U_1$ and $U_2$ as in (\ref{U1U2}),
$w^{-2}z^k=g_{12}(w,z)$ is the transition function of $\mathcal{O}(-2,k)$, i.e. its \v{C}ech cohomology class, and we work in the Laurent polynomial ring
$\bc [z,z^{-1},w,w^{-1}]$
modulo $\psi(w,z)$, the defining
polynomial of $\Sigma$.

A section exists if we can find $\chi (w^{-1},w,z^{-1},z)$ so that
$\psi(w,z)\chi$ cancels all of the negative powers of $w$ and $z$ in
$w^{-2}z^k \xi_2 (w^{-1},z^{-1})$.  So $\psi(w,z)$ acts as a type of
Toeplitz operator on $\chi$.  In fact $\chi=\chi(w^{-1},w)$ and only the
coefficients of $w^{-1}$ need to be taken care of.  This is not hard
to prove, and is most efficiently expressed in terms of an exact
sequence in \v{C}ech cohomology in which we see that $q$ is also a \v{C}ech
cocycle,
\[
0\to H^0(\Sigma,\mathcal{O}(-2,k))\to
H^1(Q,\mathcal{O}(-2-k,0))\stackrel{\psi}{\to}
H^1(Q,\mathcal{O}(-2,k));
\] 
so $\chi \in H^1(Q,\mathcal{O}(-2-k,0))$ and multiplication by $\psi(w,z)$
yields a linear map
\begin{equation} \label{eq:psi}
    \bc^{k+1}\cong H^1(Q,\mathcal{O}(-2-k,0))\stackrel{\psi}{\to}
    H^1(Q,\mathcal{O}(-2,k))\cong \bc^{k+1}
\end{equation}
with kernel corresponding to sections of
$H^0(\Sigma,\mathcal{O}(-2,k))$.  Since
$H^0(\Sigma,\mathcal{O}(-2,k))$ vanishes, the kernel is trivial and
the determinant of (\ref{eq:psi}) is nonzero.  In the tetrahedral
case (\ref{eq:psi}) is represented by the matrix
\[ 
\Psi=\left(\begin{array}{cccc}1&0&-i\alpha&0\\0&3&0&i\alpha\\ 
i\alpha&0&3&0\\0&-i\alpha&0&1 
\end{array}\right),\ \ \det\Psi=(3-\alpha^2)^2\neq 0
\]
and in the octahedral case by
\[ 
\Psi=\left(\begin{array}{ccccc}1&0&0&0&\alpha\\0&4-4\alpha&0&0&0\\
0&0&6+6\alpha&0&0\\ 
0&0&0&4-4\alpha&0\\ \alpha&0&0&0&1\end{array}\right),\ \ 
\det\Psi=96(1+\alpha)^2(1-\alpha)^3\neq 0.
\]
The nonvanishing determinant restricts $0<\alpha<\sqrt{3}$ in the 
tetrahedral case and $0<\alpha<1$ in the octahedral case, in 
agreement with the discussion in section~\ref{sec:plato}.

More significant information is obtained from
\[
0\to H^0(\Sigma,\mathcal{O}(-3,k+1))\to
H^1(Q,\mathcal{O}(-3-k,1))\stackrel{\psi_1}{\to}
H^1(Q,\mathcal{O}(-3,k+1)),
\] 
where $\psi_1$ is again a Toeplitz type operator induced from
multiplication by $\psi$.  Trivial kernel, and hence nonzero
determinant of $\Psi_1$, occurs when $m>\frac{1}{2}$, since
\[
H^0(\Sigma,L^{2m-1}(k-2,0))= H^0(\Sigma,\mathcal{O}(-3,k+1))
\]
vanishes when $0<2m-1<2m+2$.  In the tetrahedral case,
\[
\det\Psi_1=4(1-3\alpha^2)^2(3-\alpha^2)^2
\] 
and hence $\alpha\neq \frac{1}{\sqrt{3}}$ for $m>\frac{1}{2}$.  
But $\alpha$ tends to
zero as $m\to\infty$, and $\alpha$ begins at $\sqrt{3}$ when $m=0$, so by
continuity, $\alpha=\frac{1}{\sqrt{3}}$ for some value of $m$.  In fact,
\[
\alpha=\frac{1}{\sqrt{3}} \; \Leftrightarrow \;  m=\frac{1}{2},
\]
since
$H^0(\Sigma,L^{2m-1}(k-2,0))=H^0(\Sigma,\mathcal{O}(k-2,0))$ has
nontrivial sections, and thus $\Psi_1$ has nontrivial kernel, so
$\det\Psi_1=0$.

The preceeding calculation is sufficient for our purposes to get the
relation (\ref{relamA4}) once we have found a suitable path $\gamma$
on $E$ from $q$ to $p$, but we need at least one more value of
$(\alpha,m)$ to be able to find $\gamma$, as explained in 
section~\ref{sec:tetrahedral}. However, similar calculations on
$H^0(\Sigma,L^{2m-2}(k-2,0))= H^0(\Sigma,\mathcal{O}(-4,k+2))$ yield a
map $\Psi_2$ with
$\det\Psi_2=4(\alpha^2+5)^2(1-3\alpha^2)^2(\alpha^2-4\alpha+1)^2
(\alpha^2+4\alpha+1)^2$, and this enables
one to conclude that 
$\alpha=2-\sqrt{3}$ when $m=1$.  This technique applies to
yield an algebraic value of $\alpha$ for any half-integer mass $m$; Table~1 
summarises these results for $m\le \frac{3}{2}$.

In the octahedral case,
\[
\det\Psi_1=16(1+5\alpha)^2(\alpha+5)^3(3\alpha-1)^3(\alpha-1)^4
\]
and using the argument above we deduce that
\begin{equation}\label{m1on24S4}
\alpha=\frac{1}{3} \Leftrightarrow
m=\frac{1}{2}.  
\end{equation}
The next values for $\alpha$, found by computing the determinants up to
$m=\frac{3}{2}$, are displayed in Table~1.


\begin{table}
\begin{center}
    \begin{tabular}{c|cc} \hline\hline
	 & $G=A_4$  & $G=S_4$\\
     $m$ & $k=3$ & $k=4$ \\\hline
	$0$ & $\sqrt{3}$ & 1 \\
	$\frac{1}{2}$ & $\frac{1}{\sqrt{3}}$ & $\frac{1}{3}$ \\
	$1$ & $2-\sqrt{3}$ & $\frac{1}{7}$ \\
	$\frac{3}{2}$ & $\sqrt{23-4\sqrt{33}}$ & $7-4\sqrt{3} $\\
	$\infty$ & 0 & 0 \\
	\hline\hline
    \end{tabular}    
    \caption{The parameter $\alpha$ for some half-integer values of $m$.}
\end{center}
\end{table}

\section{The general mass constraints} \label{sec:genmconstr}

We start by constructing the quotients $\pi$ in (\ref{elliptquot})
explicitly, and in particular a realisation of the elliptic curves $E$,
by means of invariant theory. 
To do this, we use the procedure illustrated in Example~\ref{invsfromKlein}
to find rational invariants $\hat{v}, \hat{x}, \hat{y} \in \bc(\Sigma)^{ {A}_4}$
given in the coordinates of $\bpxp$ by
\begin{eqnarray*}
	\hat{v} & = & \frac{P_3}{P_1{}^3}:= \frac{(w+z)((wz)^2-1)}{(w-z)^3},\\
    \hat{x} & = & \frac{P_4}{P_1{}^4}:= \frac{w^4 z^4 + w^4 + z^4 +
    12 w^2 z^2 + 1}{(w-z)^4} ,\\
    \hat{y} & = & \frac{P_6}{P_1{}^6}:=
    \frac{(wz)^6-((wz)^2+1)((w+z)^4+4wz(w+z)^2+(wz)^2)+1}{(w-z)^6}.
\end{eqnarray*}
Here, $P_\ell$ denote forms of degree $(\ell,\ell)$ that are projectively
invariant under (\ref{eq:gens}). They satisfy a relation (in degree $12$)
which can be written in terms of the invariants above as
\begin{equation}\label{relationA4}
4\hat{x}^3-11 \hat{x}^2-4\hat{y}^2-14 \hat{v}^2
-27\hat{v}^4+2\hat{x}
(5+9\hat{v}^2)-3=0.
\end{equation} 	
Out of these rational functions, we produce maps 
into $\bp^2$ whose restriction to each  
$\Sigma$ has $ {A}_4$-orbits as fibres.
The images are determined by a
single relation among the invariants, which can be used to describe
the elliptic curves $\Sigma/ {A}_4$.
This procedure is easily adapted to the octahedral case, as we shall explain
below.
In practice, it is convenient to choose the coordinates of the embeddings
to obtain an image in standard Weierstra\ss\ form.

\subsection{$k=3$ tetrahedral symmetry}\label{sec:k3relat}

Using equation (\ref{tetrahedral}) to eliminate 
$\hat{v}=\frac{i}{\alpha}$, we obtain from (\ref{relationA4}) 
the relation on $\Sigma$
\[
4\hat{y}^2=4\hat{x}^3-11\hat{x}^2+\left( 
10-\frac{18}{\alpha^2}\right)
\hat{x}-3+\frac{14}{\alpha^2}-\frac{27}{\alpha^4}.
\]
So we redefine the invariants as
$x:=\hat{x}-\frac{11}{12}$, $y:=2\hat{y}$, in order to obtain
the standard plane cubic
\begin{equation}\label{WeierstrA4}
    y^2=4x^3-g_2 x-g_3=: F(x),
\end{equation}
where
\begin{equation}\label{g_23A4}
    g_2=\frac{1}{12}+\frac{18}{\alpha^2}, \qquad
    g_{3}=-\frac{1}{216}+\frac{5}{2\alpha^2}+ \frac{27}{\alpha^4}.
\end{equation}
The picture is that $\pi:=[1:x:y]$ maps $ {A}_4$-orbits in
$\Sigma$ (given by equation (\ref{tetrahedral})) into $\bp^2$, and the 
image $E$ is realised as the elliptic
curve given by (\ref{WeierstrA4}).  From this equation, we read off that
$x\circ u^{-1}=\wp$ is a Weierstra\ss\ function of $E$ and $y\circ
u^{-1}=\pm \wp'$ its derivative up to sign, where $u^{-1}:\bc
\rightarrow E$ is the well-defined inverse to a ``uniformisation map". 
We use $e_1, e_2, e_3$ to denote the zeroes of $F$,
\[
F(x)=4(x-e_1)(x-e_2)(x-e_3)
\]
with $e_1+e_2+e_3=0$.  From (\ref{g_23A4}), we compute the
$j$-invariant of $E$ to be
\[
j(\alpha)=\frac{g_2{}^3}{{g_2}^3-27{g_3}^2}=\frac{\alpha^2 (\alpha^2+216)^3}
{2^6 3^3 (\alpha^2-27)^3}.
\]
One way of
realising $E$ is as a degree 3 branched cover $[1:x\circ
\pi^{-1}]:E\rightarrow \bp^1$, with branch points $[1:e_1], [1:e_2]$
and $[1:e_3]$ and suitable branch cuts.  Clearly, two zeroes of $F$
must be nonreal and complex conjugates, while the other (say $e_2$) is
real and positive since $g_3= 4 e_1 e_2 e_3>0$.  We take ${\rm
Im}\,(e_1)>0$.  Notice that ${\rm Re}\,e_1={\rm
Re}\,e_3=-\frac{1}{2}e_2$ and ${\rm Im}\,e_3=-{\rm Im}\,e_1$.

We must define branch cuts, the sheet decomposition, a holomorphic 1-form
$\omega$ and a basis of 1-homology for $E$
before we can compute the integrals in the relation 
(\ref{reciprEA4}). 
We consider the standard holomorphic 1-form on $E$ given by 
\begin{equation}\label{omegauptosign}
\omega=\pm \frac{{\rm d}x}{\sqrt{F(x)}}.
\end{equation}
up to sign.
The conventions that we shall follow
are illustrated in Figure~2. One branch cut runs from $e_2$ to $\infty$
downwards while the other one joins $e_1$ to $e_3$ avoiding the negative
real axis; these separate $E$ into
two sheets (i) and (ii). 
The poles of $\frac{{\rm d}w}{w}-\frac{{\rm d}z}{z}$ are easily seen to be
\begin{eqnarray*}
    & p_1=(\bar{\rho}\sqrt{\alpha},0), \quad
    p_2=(-\bar{\rho}\sqrt{\alpha},0), &\\
    & q_1=(0,{\rho}\sqrt{\alpha}), \quad q_2=(0,-{\rho}\sqrt{\alpha}), &
\end{eqnarray*}
where $\rho:=e^{\pi i/4}$.
Their images on $E \subset \bp^2$ can be computed as
\begin{eqnarray*}
    &\D p:=\pi(p_1)=\left[1:\frac{1}{12}-\frac{1}{\alpha^2}:
    -\frac{2i}{\alpha}\left(1+\frac{1}{\alpha^2} \right) \right]=\pi(p_2),&\\
    &\D q:=\pi(q_1)=\left[1:\frac{1}{12}-\frac{1}{\alpha^2}:
    +\frac{2i}{\alpha}\left(1+\frac{1}{\alpha^2} \right) \right]=\pi(q_2),&
\end{eqnarray*}
and they lie on the same fibre of the double cover $[1:x] :E\rightarrow
\bp^1$, as we claimed before.
Since $\alpha<\sqrt{3}$ implies $\frac{1}{12}-\frac{1}{\alpha^2}<0$, 
the common projection of both $p$ and $q$ onto the $x$-plane lies on 
the negative real axis.
As before, paths are drawn with
continuous or dotted lines according to whether they lie on sheets 
(i) or (ii), and we have defined $a,b$ and the
path from $p$ to $q$ (avoiding $a$ and $b$) following this 
notation; in particular, $q$ lies on 
sheet (i) and $p$ on sheet (ii). 
We choose the signs in (\ref{omegauptosign})
so that the coefficient at $p$ has positive imaginary part.

\begin{figure}[ht] 
    \begin{center}
	\vspace{.8cm}
	\input{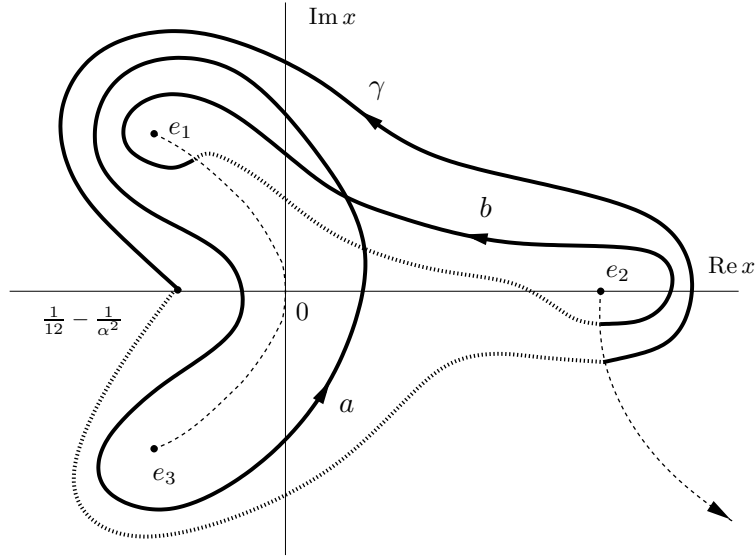}
	\small{
	\caption{Tetrahedrally symmetric 3-monopole: contours of 
             integration on $E=\Sigma/ {A}_4$.}
	}
    \end{center}
\end{figure}

Equation (\ref{reciprEA4}) relates the periods 
\begin{eqnarray}
    &\D 2 \varpi := \oint_{a}
    \omega=2\int_{e_2}^{+\infty}\frac{{\rm d}x}{\sqrt{F(x)}}& \label{hpvarpi3A4}\\
    &\D 2 \varpi' := \oint_{b}
    \omega=-2i \int_{e_2}^{e_1}
            \frac{{\rm d}x}{\sqrt{-F(x)}}& \label{hpvarpip3A4}
\end{eqnarray}
to the integral
\begin{equation}\label{intRHSA4}
    \int_{p}^{q}\omega =2i\int^{-\infty}_{\frac{1}{12}-\frac{1}{\alpha^2}}
    \frac{{\rm d}x}{\sqrt{-F(x)}}
\end{equation}
as
\begin{equation}\label{relintsA4}
    \ell_1 \varpi + \ell_2 \varpi' =
    2i(2m+3)\int^{-\infty}_{\frac{1}{12}-\frac{1}{\alpha^2}}
    \frac{{\rm d}x}{\sqrt{-F(x)}}.
\end{equation}
Notice that $\varpi$ is real, while the integral (\ref{intRHSA4}) is 
pure imaginary, and that one has    
\begin{eqnarray*}
    2 \;{\rm Re} \,\varpi' &= & -i \int_{e_2}^{e_1}\frac{{\rm d}x}{\sqrt{-F(x)}} 
    +i\,
    \overline{\int_{e_2}^{e_1}\frac{{\rm d}x}{\sqrt{-F(x)}}} \\
    &=&-i
    \int_{e_1}^{e_2}\frac{{\rm d}x}{\sqrt{-F(x)}}+i\int_{e_2}^{e_3}
    \frac{{\rm d}x}
    {\sqrt{-\bar{F}(x)}}\\
    &=&i\int_{e_1}^{e_3}\frac{{\rm d}x}{\sqrt{-F(x)}}\\
    &=&\varpi.
\end{eqnarray*}
Taking real parts in both sides of equation (\ref{relintsA4}), one
then obtains the relation
\begin{equation}\label{l1eql2}
    \ell_2 = -2 \ell_1.
\end{equation}

These integrals are most conveniently dealt with using uniformisation. 
The standard uniformisation point is the set of 2-poles of $\wp$, thus
of 2-poles of $x$:
\[
x=\infty \Leftrightarrow w=z;  
\]
an easy check gives that all the six points of $\Sigma \cap \bp^1_{\Delta}$
are poles, and they are mapped to $[0:0:1]$ by $\pi$, the branch point
at infinity of $[1:x\circ \pi^{-1}] :E\rightarrow \bp^1$.  
One has $\wp(\varpi)=e_2$ and $\wp(\varpi')=e_3$, so to conform with standard
practice we should use the half-periods
\begin{eqnarray}
&\varpi_2=\varpi, \nonumber &\\
&\D \varpi'=\frac{\varpi_1+\varpi_2}{2} \Rightarrow \varpi_1=2\varpi'-\varpi.& 
\label{hpvarpi13A4}
\end{eqnarray}
Now we can write formally
\begin{eqnarray}
i\int^{-\infty}_{\frac{1}{12}-\frac{1}{\alpha^2}}\frac{{\rm d}x}{\sqrt{-F(x)}}
&=&    \int_{\wp\circ\wp^{-1}(\frac{1}{12}-\frac{1}{\alpha^2})}^{\wp(0)}
    \frac{{\rm d}\wp}{\wp'} \nonumber \\
&=&\int_{\wp^{-1}(\frac{1}{12}-\frac{1}{\alpha^2})}^{0}{\rm d}u 
 \nonumber \\
&=&   \mp {\wp^{-1}\left(\frac{1}{12}-\frac{1}{\alpha^2}\right)},\label{wpcheat}
\end{eqnarray}
where $\wp^{-1}$ is determined only up to sign (as $\wp$ has order 2 in a
fundamental region)
and up a point in the lattice $\Lambda = 2\varpi_1\bz \oplus 2 \varpi_2 \bz$ 
inside the $u$-plane.
Equation (\ref{relintsA4}) then gives, making use of (\ref{l1eql2}),
\[
\frac{(\varpi -2 \varpi')\ell_1}{2(2m+3)}\equiv \pm
\wp^{-1}\left( \frac{1}{12}-\frac{1}{\alpha^2}\right) \quad \mod \Lambda.
\]
The ambiguity introduced in (\ref{wpcheat}) is removed once we apply
$\wp$ to both sides of this equation:
\[
\wp\left( \frac{\varpi_1 \ell_1}{2(2m+3)} \right)=
\frac{1}{12}-\frac{1}{\alpha^2},
\]
The only information missing is the integer $\ell_1$ determining 
$c=\ell_1(1,-2) \in H_1(E;\bz)$, but this
can be obtained directly from (\ref{relintsA4}) and (\ref{l1eql2}) 
using the data $(\alpha,m)$ we calculated in section~\ref{sec:halfintm}.
We always find
\[
\ell_1=4
\]
by evaluating the integrals numerically for all the 
positive values of $m$ in Table~1.
This confirms that we are working with a path $\gamma$ on $E$ that
is the image of a path with
zero intersection with the 1-cycles of $\Sigma$.
We end up with equation (\ref{relamA4}), and this completes the proof of Theorem~\ref{th:3A4}.

\subsection{$k=4$ octahedral symmetry}

When (\ref{tetrahedral4}) is used to eliminate
$\hat{x}=1-(1+i\beta\tilde{v})/\alpha$, one obtains the relation on $\Sigma$
\begin{eqnarray} \label{relation4A4}
    -4\hat{y}^2-27\hat{v}^4+\frac{2i\beta}{\alpha}\left(
\frac{2\beta^2}{\alpha^2}-9
    \right)\hat{v}^3+\left( 4-\frac{18}{\alpha} + \left(
    \frac{12}{\alpha}-1\right)\frac{\beta^2}{\alpha^2} \right)\hat{v}^2+ \nonumber \\
    \frac{2i\beta}{\alpha^2}\left( 1-\frac{6}{\alpha}\right)\hat{v} +
    \frac{1}{\alpha^2}\left(1-\frac{4}{\alpha}\right)&=&0.
\end{eqnarray}
In principle, one could follow the same procedure as in the $k=3$ case to
obtain the relation among $\alpha,\beta$ and $m$, but a calculation for general
values of the parameters is out of hand as one now must deal with the
roots of a quartic polynomial, and these then give rise to a less symmetric 
form of the Weierstra\ss\ equation~(\ref{WeierstrA4}). In the following, we shall only consider 
the case of octahedral symmetry $\beta=0$, working with $\Sigma/S_4$
rather than $\Sigma/A_4$, which considerably simplifies the calculations
as one should expect.

The extension from tetrahedral to octahedral symmetry can
be realised by adding the rotation (\ref{extraS4}),
which changes the tetrahedral invariants as
\[
\hat{v} \mapsto -\hat{v}, \quad
\hat{x} \mapsto \hat{x}, \quad
\hat{y} \mapsto  -\hat{y}. 
\]
Therefore, 
\[
\hat{x}, \quad \hat{v}\hat{y},\quad \mbox{and}\quad \hat{v}^2
\]
are octahedral invariants, and a cubic relation among them on $\Sigma$ can be 
obtained simply by multiplying both sides of (\ref{relation4A4}) by 
$\hat{v}^2$. This relation is brought to Weierstra\ss\ form 
(\ref{WeierstrA4}) if one uses the coordinates
\[
y:=\sqrt{2} i \hat{v} \hat{y},\qquad 
x:=\frac{3}{2}\hat{v}^2 + \frac{1}{3\alpha} -\frac{2}{27};
\]
the elliptic invariants are now
\begin{equation}\label{g_2S4}
g_2=\frac{16}{253}-\frac{16}{27\alpha}+\frac{5}{3\alpha^2}-
\frac{4}{3\alpha^3}
\end{equation}
and
\begin{equation}\label{g_3S4}
g_3=\frac{64}{19683}-\frac{32}{729 \alpha}+\frac{2}{9\alpha^2}
-\frac{41}{81 \alpha^3}+
\frac{4}{9 \alpha^4}.
\end{equation}
These yield the $j$-invariant
\[
j(\alpha)=\frac{(432 \alpha^3 -4048\alpha^2+11385\alpha - 9108)^3}
{2^2 3^3 11^3 23^3 (\alpha-4)^2 (\alpha-3)^3}.
\]
Again, we realise $E:=\Sigma/ {S}_4$ as the image of the map
$\pi:= [1:x:y]$ to $\bp^2$.
It is easy to see that $F(x)$ has again a real positive zero $e_2$ and two distinct complex conjugate zeroes $e_1, e_3$; we
choose the convention ${\rm Im}\,e_1>0$ as before.

The differential $\frac{{\rm d}w}{w}-\frac{{\rm d}z}{z}$ on 
$\Sigma$ has four poles 
with positive residue and four poles with negative residue, and their
images on $E \subset \bp^2$ are, respectively,
\begin{eqnarray*}
p=\left[1:-\frac{2}{27}-\frac{7}{6\alpha} :\sqrt{2}i\frac{1+\alpha}{\alpha^2}\right],\\
q=\left[1:-\frac{2}{27}-\frac{7}{6\alpha} :-\sqrt{2}i\frac{1+\alpha}{\alpha^2}\right].
\end{eqnarray*}
We should integrate on $\Sigma$ along four paths,
from $\pi^{-1}(p)$ to $\pi^{-1}(q)$ points, which do not cross a basis 
of 1-cycles of $\Sigma$. Contrary to the $k=3$ case, it will turn out that
the image of such paths will have nontrivial intersection with a standard
basis of $E$. We follow the same conventions as above
for the covering $[1:x] : E\rightarrow \bp^1$ and the periods (\ref{hpvarpi3A4})--(\ref{hpvarpip3A4}), but not for 
the path $\gamma$, which we define as having 
$\#\langle \gamma, b \rangle = +1$ as illustrated in Figure~3.

\begin{figure}[ht] 
    \begin{center}
	\vspace{.8cm}
	\input{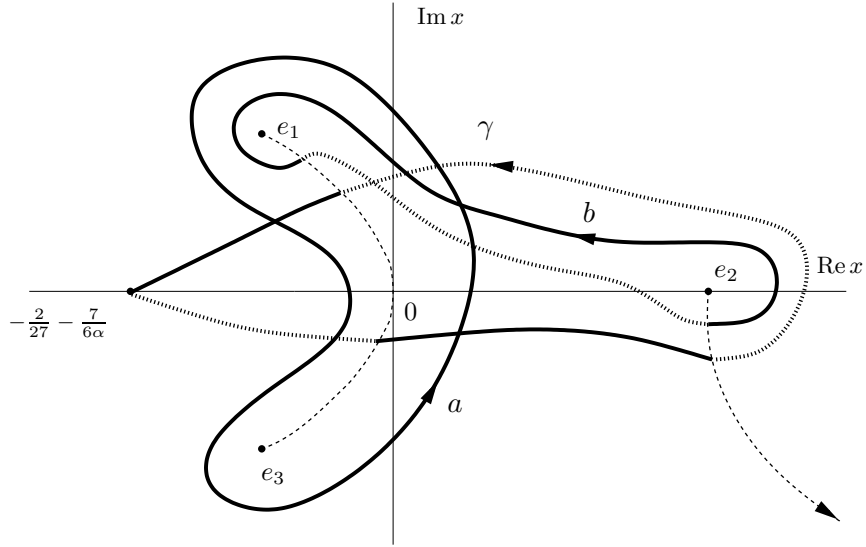}
	\small{
	\caption{Octahedrally symmetric 4-monopole: contours of 
             integration on $E=\Sigma/ {S}_4$.}
	}
    \end{center}
\end{figure}

We obtain the relation for $\omega$ and $c$ as in (\ref{omegauptosign}) and (\ref{cA4}),
\[
\oint_{c}\omega = 8 (m+2) \int_{p}^{q} \omega,
\]
and using our conventions we find
\begin{equation}\label{relintsS4}
\ell_1 \varpi + \ell_2\varpi' = 4i(m+2)\int^{-\frac{2}{27}-
\frac{7}{6\alpha}}_{e_2} \frac{{\rm d}x}{\sqrt{-F(x)}}.
\end{equation}
Again, (\ref{l1eql2}) holds, and we are led to the relation
\[
\wp\left( \frac{\varpi_1 \ell_1}{2(m+2)} \right)=
\frac{-4\alpha^4+10\alpha^3-115\alpha^2+60\alpha-3}
{54 \alpha^2 (\alpha + 1)^2},
\]
where we made use of the duplication formula for $\wp$.
Given (\ref{m1on24S4}), we find from (\ref{relintsS4})
\[
\ell_1=6,
\]
and the same result is obtained for the other $(\alpha,m)$ data, 
which confirms
that the right path $\gamma$ has been chosen. 
This completes the proof of Theorem~\ref{th:4A4}.

\section{Rational and infinite mass} \label{sec:speccases}

This final section illustrates how one may deal with the general mass constraints (\ref{relamA4}) and (\ref{relamS4}) in two important concrete 
cases, providing nontrivial checks to our calculations.

\subsection{Solutions for rational mass}

We shall now explain briefly how to solve the mass conditions (\ref{relamA4}) and (\ref{relamS4}) 
explicitly to calculate $\alpha$ for rational values of the monopole mass.
We shall be making use the fact that, for $m\in \mathbb{Q}$, the left-hand side of either of these conditions is a division value of the $\wp$-function, i.e., a complex number of the form
\[
\wp\left( \frac{2 k_1\varpi_1 + 2 k_2\varpi_2}{n}\right)
\]
where $n\in\mathbb{N}$ and $k_1,k_2 \in \mathbb{Z}$ are such that 
$(k_1,k_2)\ne (0,0) \mod n$. Classical techniques in the theory of elliptic functions~\cite{KieWA} can be used to show that division values correspond exactly to the roots of the so-called {\em special division equation} 
$P_{n}(\wp(u))=0$, where $P_n(\wp)$  is a
polynomial of degree $\frac{n^2-1}{2}$ or $\frac{n^2-4}{2}$ defined by
\[
P_{2\ell+1}(\wp(u)):=\frac{\psi_{2\ell+1}(u)}{2\ell+1} \qquad \mbox{or} \qquad
P_{2\ell+2}(\wp(u)):=-\frac{\psi_{2\ell+2}(u)}{(\ell+1)\wp'(u)}
\]
according to whether $n$ is odd or even; here,
\[
\psi_n (u):=\frac{(n-1)^{n-1}}{\left(\prod_{j=1}^{n-1}j!\right)^{2}}
\left|\begin{array}{cccc}
\wp'(u)&\wp''(u) & \cdots & \wp^{(n-1)}(u)\\
\wp''(u)& \wp'''(u) & \cdots & \wp^{(n)}(u)\\
\vdots & \vdots & \ddots & \vdots \\
\wp^{(n-1)}(u) & \wp^{(n)}(u) & \cdots & \wp^{(2n-3)}(u)   
\end{array}\right|
\]
is easily seen to satisfy $\psi_n(-u)=(-1)^{n+1}\psi_n(u)$, so indeed
$P_n(\wp) \in \mathbb{Q}[g_2,g_3][\wp]$.

To calculate $\alpha$ for $m\in \mathbb{Q}$, one 
replaces $\wp(u)$ by the right-hand side of (\ref{relamA4}) or (\ref{relamS4}) in the special division equation
for the corresponding elliptic curve $E=\Sigma/G$, thus obtaining a polynomial equation for $\alpha$; by inspection, the minimal 
polynomial for $\alpha$ is recovered
as one of the factors, and hence $\alpha$ itself. This argument shows that, for 
$m\in \mathbb{Q}$, equations (\ref{relamA4}) and (\ref{relamS4})
determine $\alpha$ as an algebraic number.

The results of Table~1 in section~\ref{sec:halfintm} are easily checked
using this procedure; in particular, we can compute 
$\lim_{m\rightarrow 0}\alpha$ and
write the limit ``nullaron" curves in the form (\ref{nullaroncurve}), from 
which we can recover the corresponding rational maps. For $m \ge 2$,
the algorithm in section~\ref{sec:halfintm} becomes impractical, whereas
the special division equation can still be used to generate minimal polynomials for the parameter $\alpha$.
In addition, one can calculate in principle the value 
of $\alpha$ for any $m\in \mathbb{Q}$.
For example, in the $k=3$, $G=A_4$ case, for $m=\frac{1}{3}$, we obtain 
the minimal polynomial
\[
\alpha^{10}-39\alpha^8 + 506\alpha^6 + 866 \alpha^4 - 715 \alpha^2 - 11,
\]
which yields $\alpha\simeq 0.791875$ as the unique positive real root larger than $\alpha|_{m=\frac{1}{2}}=\frac{1}{\sqrt{3}}$.

\subsection{Euclidean limit}

As explained in section~\ref{sec:euclim2}, centred
spectral curves degenerate to $k$ copies of
the diagonal $\bp^1_\Delta$ when the
limit $m\rightarrow \infty$ is taken, but a rescaling of the metric
yields curves in $T\bp^1$ which can be interpreted as spectral curves of
euclidean monopoles. This limit process allows one to
derive the spectral curves of platonic monopoles in euclidean space from
our results for hyperbolic monopoles; it will also provide a 
nontrivial check on our calculations.

We focus on the case $k=3$, $G=A_4$ for brevity. Given (\ref{eq:geodlim}),
we expect the rescaled parameter $\tilde{\alpha}:=m^3 \alpha$ to tend to
a finite limit as $m\rightarrow \infty$ (and $\alpha \rightarrow 0$),
and our aim is to calculate this limit using the relation (\ref{relamA4}). 
The natural way to proceed is
to rescale the $A_4$-invariants on $\Sigma$ in such a way that 
the quotient elliptic curve $E=\Sigma/A_4 \subset \bp^2$ has a 
sensible limit; thus we use $\tilde{x} := x/m^4$ and $\tilde{y} := y/m^6$
to embed $E$ in $\bp^2$.
When $m\rightarrow \infty$, one obtains the Weierstra\ss\ equation 
for $E$ 
\[
\tilde{y}^2=4\tilde{x}^3-\frac{27}{\tilde{\alpha}^2}=:
4(\tilde{x}-\tilde{e}_1)(\tilde{x}-\tilde{e}_2)(\tilde{x}-\tilde{e}_3)
=:\tilde{F}(\tilde{x}),
\]
where we use conventions for the invariants (and periods) consistent with 
section~\ref{sec:k3relat}. This is the same elliptic curve as in the
calculations in section~9 of~\cite{HitManMur}, since the $j$-invariant is
zero in both cases.
Clearly, the periods transform as $\tilde{\varpi}_{j} = m^{2} \varpi_j$
under our rescaling.
Using the asymptotics $\wp(\rho)=\frac{1}{\rho^2}+O(\rho^2)$ as 
$\rho \rightarrow 0$, one obtains from (\ref{relamA4}) the relation
\begin{equation}\label{relatlim}
\frac{m^6}{\tilde{\varpi}_1^2}+O(\frac{1}{m^6})=\frac{1}{12}-\frac{m^6}{\tilde{\alpha}^2} \quad
\stackrel{m\rightarrow \infty}{\Longrightarrow} \quad
\tilde{\varpi}_1=i \tilde{\alpha}
\end{equation}
for the limit elliptic curve. Now we evaluate, in the limit,
\begin{eqnarray}
\tilde{\varpi}_{1}&=&-2i \int_{\tilde{e}_2}^{\tilde{e}_1}
\frac{{\rm d}\tilde{x}}{\sqrt{-\tilde{F}(\tilde{x})}}-
\int_{\tilde{e}_2}^{+\infty}
\frac{{\rm d}\tilde{x}}{\sqrt{\tilde{F}(\tilde{x})}} \nonumber\\
&=&-\frac{2^{1/3}e^{2\pi i/3}\sqrt{\pi}\tilde{a}^{2/3}
\Gamma\left(\frac{1}{3}\right)}{3 \Gamma\left( \frac{5}{6}\right)}
-\frac{2^{1/3}\sqrt{\pi}\tilde{a}^{2/3}
\Gamma\left(\frac{1}{6}\right)}{9\sqrt{3}\Gamma\left( \frac{5}{3}\right)}.
\label{periodlim}
\end{eqnarray}
Equating the right-hand sides of (\ref{relatlim}) and (\ref{periodlim}), we find
\[
\tilde{\alpha}=\frac{\Gamma\left(\frac{1}{3}\right)^9 }{2^6 \pi^3};
\]
thus the limit curve in $T\bp^1$
\[
\eta^{3}+2i \tilde{\alpha}\zeta(\zeta^4-1)=0
\]
is precisely the spectral curve of a euclidean 3-monopole with tetrahedral
symmetry as found in~\cite{HitManMur,HouManRom}.

\vspace{.75cm}

\noindent
{\large \bf Acknowledgements}\\[5pt]
The authors are grateful to Michael Murray for discussions and support.
The second author's work is financed by the Australian Research Council,
and he would like to thank the Department of Mathematics and
Statistics of the University of Melbourne for hospitality.

\vspace{.5cm}

\begin{small}

\newcommand{\href}[1]{}
\newcommand{\hpeprint}[1]{\href{http://arXiv.org/abs/#1}{\texttt{#1}}}

\end{small}

\end{document}